\documentclass[12pt,preprint]{aastex}

\def\pf{p_f}
\def\pr{p_r}

\def\rf{r_f}
\def\rr{r_r}

\def\Gb{\Gamma}
\def\Gej{\Gamma_{\rm ej}}

\def\Lej{L_{\rm ej}}
\def\vej{v_{\rm ej}}

\def\rhoej{\rho_{\rm ej}}
\def\rhoejRS{\rho_{\rm ej}{\rm (RS)}}
\def\nejRS{n_{\rm ej}{\rm (RS)}}
\def\GejRS{\Gamma_{\rm ej}{\rm (RS)}}

\def\tobs{t_{\rm obs}}
\def\tlab{t_{\rm lab}}
\def\bargp{\bar \gamma_p}
\def\gm{\gamma_m}
\def\gc{\gamma_c}
\def\nuflu{\nu_{\rm fluid}}
\def\nulab{\nu_{\rm lab}}
\def\nuobs{\nu_{\rm obs}}

\def\be{\begin{equation}}
\def\ee{\end{equation}}
\def\beq{\begin{eqnarray}}
\def\eeq{\end{eqnarray}}

\begin{document}

\title{Dynamics and Afterglow Light Curves of GRB Blast Waves \\
with a Long-lived Reverse Shock}

\author{Z. Lucas Uhm\altaffilmark{\dag}, Bing Zhang}
\affil{Department of Physics and Astronomy, University of Nevada - Las Vegas, \\
4505 S. Maryland Parkway, Las Vegas, NV 89154, US}

\author{Romain Hasco\"et, Fr\'ed\'eric Daigne, Robert Mochkovitch}
\affil{Institut d'Astrophysique de Paris, UMR 7095 Universit\'e Pierre et Marie Curie - CNRS, \\
98 bis boulevard Arago, Paris 75014, France}

\author{Il H. Park}
\affil{Department of Physics, Sungkyunkwan University, Suwon 440-746, Korea}

\altaffiltext{\dag}{E-mail: uhm@physics.unlv.edu}

\begin{abstract}
We perform a detailed study on the dynamics of a relativistic blast wave 
with the presence of a long-lived reverse shock (RS). 
Although a short-lived RS has been widely considered, 
the RS is believed to be long-lived as a consequence of a stratification 
expected on the ejecta Lorentz factors.
The existence of a long-lived RS makes the forward shock (FS) dynamics 
to deviate from a self-similar Blandford-McKee solution. 
Employing the ``mechanical model'' that correctly incorporates 
the energy conservation, 
we present an accurate solution for both the FS and RS dynamics. 
We conduct a sophisticated calculation of the afterglow emission. 
Adopting a Lagrangian description of the blast wave, we keep track of 
an adiabatic evolution of numerous shells between the FS and RS. 
An evolution of the electron spectrum is also followed individually for every shell. 
We then find the FS and RS light curves by integrating over 
the entire FS and RS shocked regions, respectively. 
Exploring a total of 20 different ejecta stratifications, 
we explain in detail how a stratified ejecta affects 
its blast wave dynamics and afterglow light curves. We show that, 
while the FS light curves are not sensitive to the ejecta stratifications, 
the RS light curves exhibit much richer features, including steep declines, 
plateaus, bumps, re-brightenings, and a variety of temporal decay indices. 
These distinctive RS features may be observable 
if the RS has higher values of the micophysics parameters than the FS. 
We discuss possible applications of our results in understanding the
GRB afterglow data.  
\end{abstract}


%--------------------------------------------------------
%
% Section 1 Introduction 
%
%--------------------------------------------------------

\section{Introduction} \label{section:introduction}

The central engine of a gamma-ray burst (GRB) ejects a relativistic 
outflow (called an ejecta) with high Lorentz factors. As the ejecta 
interacts with a surrounding ambient medium, a relativistic blast wave 
develops. The blast wave consists of two shock 
waves: the forward shock (FS) wave sweeping up the ambient medium and 
the reverse shock (RS) wave propagating through the ejecta. 
The shocked ambient medium is separated from the shocked ejecta by a 
contact discontinuity (CD), 
and a compressed hot gas between the FS and RS is called a ``blast''.

Without an extended activity of the central engine, 
the RS is expected to be short-lived if the ejecta is assumed 
to have a constant Lorentz factor $\Gej$. 
The RS vanishes as it crosses the end of the ejecta. 
The blast wave then enters a self-similar stage where the FS dynamics 
is described by the solution of Blandford \& McKee (1976) (hereafter 
BM76). This FS emission has been believed to be the main source of the 
long-lasting, broad-band afterglows (M\'esz\'aros \& Rees 1997;
Sari et al. 1998). The short-lived RS emission would be then important 
only briefly in the early afterglow phase. Thus, it was proposed 
to explain a brief optical flash detected in some GRBs
(M\'esz\'aros \& Rees 1997,1999; Sari \& Piran 1999a,b). 
The dynamical evolution of such a short-lived RS with a constant 
$\Gej$ was studied analytically (Sari \& Piran 1995; Kobayashi 2000),
under the assumption of an equality of pressure across the blast wave. 

However, a general view on the structure of the ejecta should include the
possibility that the ejecta emerges with a range of the Lorentz
factors. The shells with lower Lorentz factors gradually ``catch
up'' with the blast wave as it decelerates to a comparable Lorentz
factor. Therefore, the RS wave is believed to be long-lived in
general.  An example with a long-lived RS, where a power-law ejecta
interacts with a power-law ambient medium, was studied analytically by
assuming a constant ratio of two pressures at the FS and RS (Rees \&
M\'esz\'aros 1998; Sari \& M\'esz\'aros 2000). 
%The impact of a
%trailing shell onto the blastwave has been analytically studied
%(e.g. Kumar \& Piran 2000; Zhang \& M\'esz\'aros 2002). 

The structure or stratification of the ejecta and ambient medium could
be in fact even more general. There is no reason why it should take only 
a constant or power-law profile. Uhm (2011) (hereafter U11)
presented a semi-analytic formulation for this class of general
problems where the ejecta and ambient medium can have an arbitrary
radial stratification.  U11 takes into account a radial spread-out and
spherical expansion of such a stratified ejecta and finds which shell
of this evolved ejecta gets passed by the RS at a certain time
and radius.  U11 then finds the dynamics of the blast wave with a
long-lived RS, by employing two different methods: (1) an equality of
pressure across the blast wave (mentioned above) and (2) the
``mechanical model'' (Beloborodov \& Uhm 2006).  U11 shows that the
two methods yield significantly different dynamical evolutions and
demonstrates that the method (1) does not satisfy the energy
conservation law for an adiabatic blast wave while the method (2)
does.  The mechanical model does not assume either an equality of
pressure across the blast wave or a constant ratio of two pressures at
the FS and RS.  It shows that the ratio of two pressures should in
fact evolve in time as the blast wave propagates.

Besides these theoretical considerations, recent early afterglow
observations led by {\it Swift} revealed a perplexing
picture regarding the origin of GRB afterglows. In contrast of a
simple power-law decay feature as expected from the standard afterglow
theory, the X-ray data show more complicated features including
initial rapid declines, plateaus, and flares (e.g. Tagaliaferri et al.
2005; Burrows et al. 2005; Nousek et al. 2006; O'Brien et al. 2006; 
Chincarini et al. 2007) that reveal rich physics in the early afterglow 
phase (Zhang et al. 2006; Zhang 2007). More puzzlingly, some GRBs show 
clear chromatic behaviors of the X-ray and optical afterglows (e.g. 
Panaitescu et al. 2006; Liang et al. 2007, 2008). It is now evident 
that the FS alone cannot interpret the broad-band afterglow data for 
the majority of GRBs. 
%and there exist two or even more emission sites 
%that contribute simultaneously to the observed afterglow flux. 

%Late central engine activities have been invoked to interpret the
%unexpected afterglow features (e.g. Zhang et al. 2006; Ghisellini et al. 2007). 

Uhm \& Beloborodov (2007) and Genet et al. (2007)
independently showed that the RS-dominated afterglow flux could 
reproduce some observed afterglow features, given the assumption
that the FS emission is suppressed.
In this paper, we study in great detail the dynamics and afterglow light curves of
GRB blast waves with a long-lived RS. The purpose
is to investigate how different ejecta stratifications 
affect their blast wave dynamics and afterglow light curves. 
We explore various types of the ejecta stratification and unveil that 
there exists a whole new class of the blast wave dynamics with a rapid and 
strong evolution of the RS strength. 
In order to find an accurate solution for both the FS and RS dynamics, 
we make use of U11 with the mechanical model. As explained above,
this allows for the blast wave with a long-lived RS to satisfy the energy conservation, 
by introducing a pressure-gradient across the blast wave region. 

We perform a sophisticated calculation of afterglow emission, 
invoking a Lagrangian description of the blast wave. 
In the widely-used analytical afterglow model (e.g. Sari et al. 1998), 
it is assumed that the entire shocked material forms a single zone
with same energy density and magnetic field. The electron energy
distribution is solved only in the energy space, with no consideration
of spatial distribution within the shocked region. Beloborodov (2005) 
(hereafter B05) described a more sophisticated Lagrangian method,  
in which the postshock region is 
resolved into subshells using a Lagrangian mass coordinate. 
B05 studies an evolution of the magnetic field and power-law spectrum 
of electrons for each subshell as the blast wave propagates. 
However, the postshock material in B05 is not resolved in radius 
and all the subshells are located at the same radius. 
Also, the pressure and energy density in B05 are assumed to be constant 
throughout the postshock material. 
Improving on B05, here we have a spatial resolution into the blast region, 
allowing for our Lagrangian shells to have their own radius. 
We further introduce a pressure profile that smoothly varies over the blast.
This is because the pressure at the FS differs from the pressure at 
the RS, as discussed above. As the blast wave propagates, we keep 
track of an evolution of the pressure, 
energy density, and adiabatic index of every shell on the blast. 
We also keep track of an evolution of the magnetic field and 
power-law spectra of electrons of all shells.

Finally, in order to calculate synchrotron radiation from a spherical shell on 
the blast, we analytically find an observed spectral flux for a distant 
observer, taking into account the effects of the shell's radial velocity 
and spherical curvature. We integrate this flux over the entire blast 
to find the sum of emissions from all the shells between the FS and RS. 

In Section~\ref{section:dynamics}, we briefly summarize how we find the dynamics 
of a blast wave with a long-lived RS. 
In Section~\ref{section:light_curves}, we describe in detail 
our method of calculating afterglow light curves. 
Numerical examples are presented in Section~\ref{section:numerical_examples}, 
which exhibit various features on the blast wave dynamics and afterglow light curves. 
Our results are summarized in Section~\ref{section:discussion} (Discussion) and 
Section~\ref{section:conclusion} (Conclusion).

%--------------------------------------------------------
%
% Section 2  Dynamics
%
%--------------------------------------------------------

\section{Dynamics of a blast wave with a long-lived RS} \label{section:dynamics}

\begin{figure}
\begin{center}
\includegraphics[width=10cm]{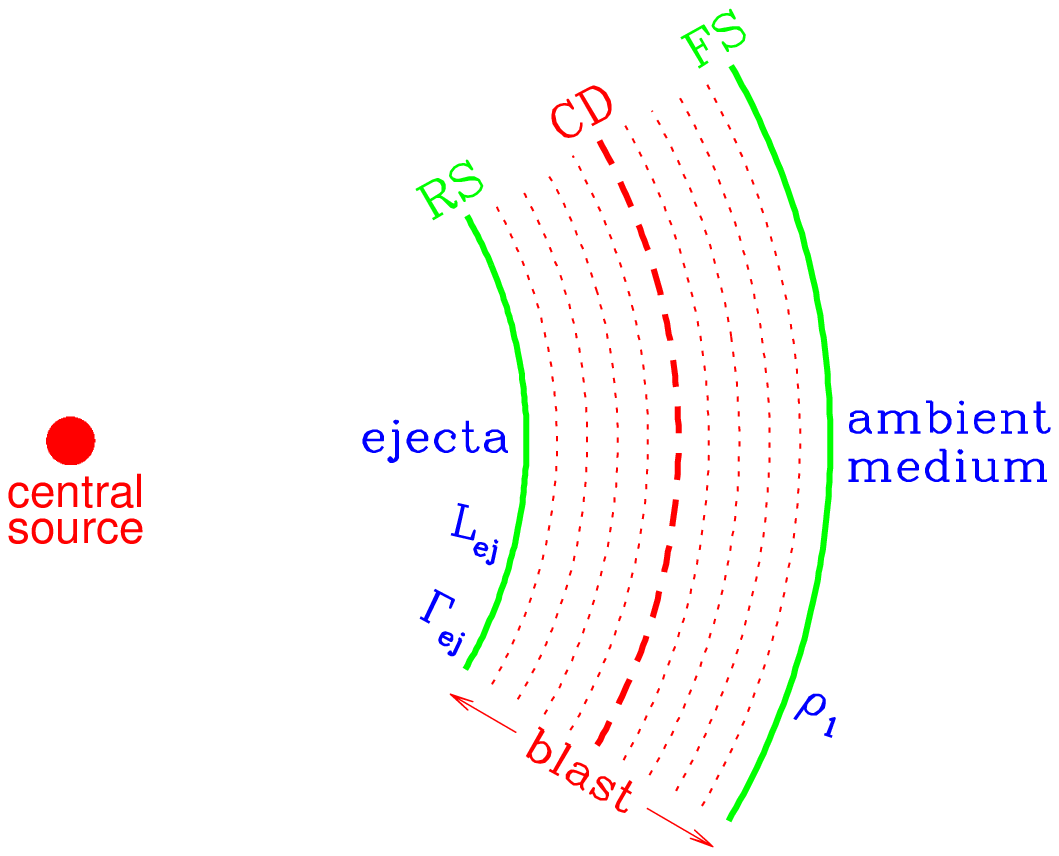}
\caption{
Illustrative diagram of a spherical blast wave. 
A forward shock (FS) wave sweeps up the surrounding ambient medium, 
and a reverse shock (RS) wave propagates through the ejecta. 
The shocked ambient medium is separated from the 
shocked ejecta by a contact discontinuity (CD). 
A Lagrangian description is employed to track an adiabatic 
evolution of all shells on the blast between the FS and RS.
} 
\label{fig:shells}
\end{center}      
\end{figure}

A schematic diagram of a spherical blast wave is shown in Figure~\ref{fig:shells}.
As mentioned in Section~\ref{section:introduction}, 
the RS wave is expected to be long-lived with a stratification on the ejecta shells. 
As the blast wave with a long-lived RS is not in the self-similar stage of BM76, 
its deceleration deviates from the solution of BM76.

In order to find such dynamics of a blast wave with a long-lived RS, 
we make use of the semi-analytic formulation presented in U11. 
The formulation has three input functions $\rho_1(r)$, $\Lej(\tau)$, and $\Gej(\tau)$,
which define the initial setup of the blast wave (see Figure~\ref{fig:shells}).
The ambient medium density $\rho_1(r)$ is allowed to take an arbitrary radial profile, 
where $r$ is the radius measured from the central engine.
The ejecta is completely specified by two other functions, i.e., 
its kinetic luminosity $\Lej(\tau)$ and Lorentz factor $\Gej(\tau)$. 
Here $\tau$ indicates an ejection time of the ejecta shells.

When two functions $\Lej(\tau)$ and $\Gej(\tau)$ are known, 
a continuity equation, which governs a radial spread-out and spherical expansion 
of a stratified ejecta, can be solved. This then yields an analytic solution 
for the ejecta density $\rhoej$ of any $\tau$-shell at radius $r$ (U11, Section 3.1), 
\be
\label{eq:ejecta_density}
\rhoej(\tau, r)=
\frac{\Lej(\tau)}{4 \pi r^2 \vej\, \Gej^2 c^2} 
\left[1-\frac{r}{c}\, \frac{\Gej^{\prime}}{(\Gej^2-1)^{3/2}}\right]^{-1}.
\ee
Here $c$ is the speed of light, and $\vej(\tau) = c\, (1-1/\Gej^2)^{1/2}$ 
is the velocity of $\tau$-shell. 
Equation (\ref{eq:ejecta_density}) is exact for a non-increasing profile 
of $\Gej(\tau)$; $\Gej^\prime(\tau) \equiv d\Gej/d\tau \leq 0$.

We then self-consistently find the path of the RS wave through 
this evolved ejecta (U11, Section 3.3). 
When the RS wave is located at radius $\rr(t)$ at time $t$, 
we numerically determine which $\tau$-shell gets shocked by the RS, and name it 
as $\tau_r(t)$-shell; the subscript $r$ in $\rr$ and $\tau_r$ refers to the RS.
In other words, the RS wave sweeps up the $\tau_r(t)$-shell at time $t$, 
which has the Lorentz factor $\Gej(\tau_r) \equiv \GejRS$ and 
the density $\rhoej (\tau_r, \rr) \equiv \rhoejRS$.

The formulation finds a dynamical evolution of the blast wave using 
the ``mechanical model'' (Beloborodov \& Uhm 2006; U11, Section 4). 
The mechanical model was developed for a relativistic blast wave 
by applying the conservation laws of energy-momentum tensor and mass flux 
on the blast between the FS and RS. Specifically, we numerically solve 
a set of coupled differential equations (78)-(80) and (92) of U11, 
which makes use of the FS and RS jump conditions (U11, Section 3.2).

The FS and RS jump conditions in U11 are derived adopting a realistic 
equation of state (EOS) with a variable index $\kappa$,
\be
\label{eq:EOS1}
p = \kappa\, (e-\rho c^2),
\qquad
\kappa = \frac{1}{3} \left(1+\frac{1}{\bar \gamma} \right),
\ee
where $p$, $e$, and $\rho$ are 
pressure, energy density, and mass density of the shocked gas, respectively, 
and $\bar \gamma$ is the mean Lorentz factor of gas particles. 
The quantity $\kappa$ smoothly varies between $2/3$ (for a non-relativistic gas) 
and $1/3$ (for an ultra-relativistic gas) as the gas temperature varies. 
This EOS differs from the exact EOS of a relativistic ideal gas (e.g., Synge 1957) 
by less than 5 \% (U11, Section 2.2). 
Since $e = \bar \gamma \, \rho c^2$, 
the EOS in Equation (\ref{eq:EOS1}) is the same as 
\be
\label{eq:EOS2}
\frac{p}{\rho c^2} = \frac{1}{3}\, \left( \frac{e}{\rho c^2}-\frac{\rho c^2}{e} \right),
\ee
which was previously introduced by Mathews (1971) 
considering a relativistic ``monoenergetic'' gas, 
and later adopted by Meliani et al. (2004) and Mignone \& McKinney (2007) 
in their numerical simulations.

%--------------------------------------------------------
%
% Section 3  Light curves
%
%--------------------------------------------------------

\section{Light curves from a blast wave with a long-lived RS} \label{section:light_curves}

Numerically solving for the dynamics of a blast wave with a long-lived RS, 
we discretize the ambient medium and ejecta into spherical mass 
shells $\delta m$. At every calculation step, a pair of shells is impulsively 
heated by shock fronts; one shell by the FS and the other by the RS. 
We follow these shells subsequently and use them as our Lagrangian shells for the blast. 
Thus, the blast is viewed as being made of many different hot shells 
that pile up from the FS and RS, as depicted in Figure~\ref{fig:shells} 
with dotted (red) curves.

Assuming an adiabatic blast wave, we keep track of an adiabatic evolution of these shells; 
an evolution of the thermodynamic quantities of shocked gas (pressure, energy density, 
adiabatic index, etc) is followed individually for every shell. 
This in turn yields an evolution of the magnetic field for the shell 
(Section~\ref{section:adiabatic_blast}). 
We also keep track of an evolution of the power-law spectrum of electrons; 
a radiative and adiabatic cooling of the spectrum is followed for every shell 
(Section~\ref{section:electron_spectrum}).

Zooming in on the blast shown in Figure~\ref{fig:shells}, 
we focus on a single spherical shell of radius $r$ expanding 
with a Lorentz factor $\Gamma$ (see Figure~\ref{fig:curvature}).
While taking into account the effects of the shell's Doppler boosting and 
spherical curvature on the synchrotron photons emitted along an observer's line of sight, 
we analytically find an observed spectral flux in terms of an observed frequency $\nuobs$ 
and observer time $\tobs$ (Section~\ref{section:curvature_effect}). 
As the blast is made of many Lagrangian shells, 
this spectral flux is summed over the blast; 
a sum of emissions from all the shells in the shocked ambient medium (or the shocked ejecta) 
is denoted by ``FS emission'' (or ``RS emission'').

\subsection{Adiabatic evolution of the blast} \label{section:adiabatic_blast}

Here we follow the adiabatic evolution of the shocked gas on a shell, 
in order to find the evolution of the magnetic field for the shell. 
An adiabatic process of a relativistic gas whose EOS is specified by 
Equation (\ref{eq:EOS1}) or (\ref{eq:EOS2}) is described by 
(U11, Section 3.4.1) 
\be
\label{eq:pm_kappa}
\frac{p}{p_m(\kappa)} = C = \mbox{const.}
\qquad
\mbox{where} 
\qquad
p_m(\kappa)
\equiv
\frac{\kappa^{5/2}~(\frac{2}{3}-\kappa)^{5/2}}
{(\kappa-\frac{1}{3})^4}.
\ee 
The function $p_m(\kappa)$ is monotonically decreasing in its valid range, 
$\frac{1}{3} < \kappa < \frac{2}{3}$.

Let us consider a shell $\delta m^i$ either in the ambient medium or in the unshocked ejecta; 
an index $i$ is added to specify this shell. 
When the shell $\delta m^i$ is shocked by the FS (or the RS), 
the jump conditions of the FS (or the RS) determine its initial thermodynamic 
quantities: pressure $p_0^i$, energy density $e_0^i$, mass density $\rho_0^i$, 
and quantity $\kappa_0^i$ (U11, Section 3.2). 
Substituting the initial values $p_0^i$ and $\kappa_0^i$ into Equation (\ref{eq:pm_kappa}), 
we determine the constant $C^i$ of the shell $\delta m^i$,
\be
\label{eq:const_C}
\frac{p^i}{p_m(\kappa^i)} = C^i = \frac{p_0^i}{p_m(\kappa_0^i)}.
\ee 
Thus, if we know $p^i$ or $\kappa^i$ of the shell $\delta m^i$ at later times, 
we can subsequently follow an adiabatic evolution of the shell.

Solving the mechanical model, at every calculation step, say at time $t$, 
we know $\rf(t)$, $\rr(t)$, $\pf(t)$, $\pr(t)$, and $P(t)$, which are 
radii of the FS and RS, pressures at the FS and RS, and 
integrated pressure over the blast, respectively. 
Thus, an instantaneous pressure profile for the blast ($\rr <r< \rf$)
may be approximated by a quadratic function $p(r)$, which 
(1) matches two boundary values (i.e., $p(\rf(t))=\pf(t)$ and $p(\rr(t))=\pr(t)$) 
and (2) satisfies the integrated pressure $P(t)=\int_{\rr}^{\rf} p(r)\, dr$ 
(U11, Section 5).

As we also know the radius $r^i$ of the shell $\delta m^i$ at time $t$, 
we now have the pressure $p^i$ of the shell $\delta m^i$ at time $t$: $p^i(t) = p(r^i(t))$. 
Equation (\ref{eq:const_C}) then allows us to numerically find 
the quantity $\kappa^i$ of the shell $\delta m^i$ at time $t$. 
All other thermodynamic quantities of the shell can be found accordingly. 
For instance, Equation (\ref{eq:EOS1}) gives the thermal energy 
density $e_{\rm th}^i$ of the shell as 
\be
e_{\rm th}^i(t) \equiv e^i(t) - \rho^i(t)\, c^2 = \frac{p^i(t)}{\kappa^i(t)}. 
\ee
The electrons in the shell $\delta m^i$ emit synchrotron radiation 
in a magnetic field $B^i$. The field is unknown and parameterized 
by a micophysics parameter $\epsilon_B$, which is the ratio of 
field energy density to thermal energy density\footnote{
See Ioka et al. (2006) for a discussion on a possible time dependent 
evolution of the microphysics parameters.}: 
$\epsilon_B = (\frac{1}{8\pi} B^{i\, 2})/e_{\rm th}^i$. 
Thus, the magnetic field $B^i$ of the shell $\delta m^i$ at time $t$ 
is given by
\be
B^i(t) 
= \left[ 8\pi \epsilon_B \, e_{\rm th}^i(t) \right]^{1/2} 
= \left[ 8\pi \epsilon_B \, \frac{p^i(t)}{\kappa^i(t)} \right]^{1/2}. 
\ee
Hereafter, we will omit the index $i$ to simplify our notation.

%--------------------------------------------------------

\subsection{Power-law spectrum of electrons} \label{section:electron_spectrum}

We assume that a non-thermal electron spectrum is created 
in a fresh shell $\delta m$ at a shock front (FS or RS); 
i.e., the electrons are accelerated into a power-law distribution 
above a minimum Lorentz factor $\gm$, 
\be
f(\gamma_e) \equiv \frac{dN}{d \gamma_e}=
K\, (\gamma_e-1)^{-p} 
\qquad \mbox{for} \qquad
\gamma_e \geq \gm,
\ee
where $K$ is a constant, $p$ is the slope of the spectrum, 
and $\gamma_e$ is the Lorentz factor of the
accelerated electrons in the fluid frame. 
The total number of electrons within the spectrum is given by   
\be
\label{eq:deltaN}
\delta N = \int_{\gm}^{\infty} f(\gamma_e)\, d \gamma_e =
\frac{K}{p-1}\, (\gm-1)^{1-p}.
\ee
The thermal energy of all electrons in the spectrum is found as
\be
\delta E 
= \int_{\gm}^{\infty} \left[(\gamma_e-1) m_e c^2 \right]\, 
  f(\gamma_e)\, d \gamma_e 
= \frac{K m_e c^2}{p-2}\, (\gm-1)^{2-p}, 
\qquad p>2,
\ee
where $m_e$ is the electron mass. 
Thus we find the mean thermal energy per electron,
\be
\label{eq:mean_thermal}
\frac{\delta E}{\delta N} = \frac{p-1}{p-2}\, (\gm-1)\, m_e c^2. 
\ee
For neutral plasma without pair-loading, if  
(1) all electrons passing through the shock become 
non-thermal\footnote{This is a usual assumption 
in the afterglow literature, although it is not necessarily true. 
See Genet et al. (2007) for examples of afterglow light curves 
in the long-lived RS scenario, obtained by assuming 
that only a small fraction $\zeta$ of the electrons become non-thermal.
}, and 
(2) a fraction $\epsilon_e$ of the shock energy goes 
to the electrons\footnote{The dominant fraction of shock energy goes to the protons, 
which dominate the pressure of the blast and evolve adiabatically. 
We may estimate the pressure $p_e$ of the electrons prescribed 
by the fraction $\epsilon_e$. For relativistic electrons,
Equation (\ref{eq:EOS1}) gives $p_e = (1/3)\, e_{{\rm th}, e}$ 
where $e_{{\rm th}, e}$ is the thermal energy density of electrons.
Equations (\ref{eq:mean_thermal}) and (\ref{eq:mean_thermal2}) yield 
$e_{{\rm th}, e} = 
n\, (p-1)/(p-2)\, \gm\, m_e c^2 =
n\, \epsilon_e\, (\bargp - 1)\, m_p c^2$
with the number density $n$ of electrons or protons. 
Thus we get
$p_e = (n/3)\, \epsilon_e\, (\bargp - 1)\, m_p c^2$.
It can be compared to the proton pressure $p_p$, 
for which we use Equation (19) of U11; 
$p_p = (n/3)\, (\bargp^2-1)/\bargp\, m_p c^2$.
This then gives the ratio
$p_e/p_p = \epsilon_e\, \bargp/(\bargp+1)$.
}, 
then the mean thermal energy per electron is alternatively given by 
\be
\label{eq:mean_thermal2}
\frac{\delta E}{\delta N} =
\epsilon_e\, (\bargp - 1)\, m_p c^2.
\ee
Here $m_p$ is the proton mass and $\bargp$ is the mean Lorentz factor 
of the protons in the postshock medium. 
Equations (\ref{eq:mean_thermal}) and (\ref{eq:mean_thermal2}) are 
combined to yield the lowest Lorentz factor $\gm$ as 
\be
\gm = 1+\frac{p-2}{p-1}\, \frac{m_p}{m_e}\, \epsilon_e\, (\bargp - 1).
\ee
For a fresh shell created at a shock front, 
$\bargp$ equals the shock strength, i.e., the relative Lorentz factor 
of the preshock to the postshock medium (BM76; U11, Equation (14)).
Thus, for the fresh shell at the FS, $\bargp$ equals $\Gb$, 
the Lorentz factor of the blast, as the ambient medium is at rest in the lab. frame.
For the fresh shell at the RS, $\bargp$ equals $\gamma_{43}$ given by 
\be
\label{eq:g43}
\gamma_{43} = \Gb \Gej - \left[ (\Gb^2-1)(\Gej^2-1) \right]^{1/2},
\ee
where $\Gej$ is to be evaluated for the $\tau_r$-shell; $\Gej = \GejRS$.

We use Equation (\ref{eq:deltaN}) to find the constant $K$ 
and substitute it into the electron spectrum,
\be
f(\gamma_e)=\delta N\, \frac{p-1}{\gm-1} 
            \left[ \frac{\gamma_e-1}{\gm-1} \right]^{-p}.
\ee
This spectrum created at the shell $\delta m$ evolves 
as the blast wave propagates. We track the adiabatic and radiative 
cooling of the electron spectrum as follows.

1. Adiabatic cooling of $\gm$: an adiabatic cooling 
of relativistic electrons is described 
as $\gamma_e \propto p^{1/4}$. 
Thus the minimum Lorentz factor $\gm$ also evolves 
as $\gm \propto p^{1/4}$. 
Here $p$ is the pressure of the shell, 
not the slope of the electron spectrum.

2. Radiative and adiabatic cooling at high $\gamma_e$: 
the electrons at high $\gamma_e$ in the spectrum 
experience a radiative and adiabatic cooling, 
which is described by the first and the second term below, respectively,
\be
\label{eq:gamma_e}
\dot \gamma_e = -\frac{1}{6\pi}\, \frac{\sigma_T}{m_e c}\, B^2\,(1+Y)\, \gamma_e^2 
                + \frac{1}{4}\, \frac{\dot p}{p}\, \gamma_e.
\ee
Here $\sigma_T$ is the Thomson cross section, 
and the dot indicates a derivative with respect to $t^{\prime}$, 
the time measured in the co-moving fluid frame. 
The Compton parameter $Y$ describes a relative contribution of 
inverse Compton scattering to the cooling rate of electrons.

Equation (\ref{eq:gamma_e}) defines a cutoff Lorentz factor $\gc$ 
at the high end of the spectrum. 
Dividing Equation (\ref{eq:gamma_e}) by $-\gamma_e^2$, we get 
\be
\label{eq:gamma_c}
\frac{d}{d t^{\prime}} \left(\frac{1}{\gc}\right) = 
\frac{1}{6\pi}\, \frac{\sigma_T}{m_e c}\, B^2\,(1+Y) - 
\frac{1}{4} \left(\frac{1}{\gc}\right) \frac{d \ln p}{d t^{\prime}} .
\ee
The parameter $Y$ can be evaluated as (Sari \& Esin 2001)
\be
(1+Y)Y = \frac{\epsilon_e}{\epsilon_B}\,\eta 
\qquad \mbox{or} \qquad
Y = -\frac{1}{2}+\sqrt{\frac{1}{4}+\frac{\epsilon_e}{\epsilon_B}\,\eta},
\ee
where
\beq
\label{eq:eta}
\eta =
\left\{ 
\begin{array}{ll}
\left( \gm/\gc \right)^{p-2} \qquad & \mbox{for} \quad \gm \leq \gc, \\
\quad 1                      \qquad & \mbox{for} \quad \gc < \gm.
\end{array}
\right.
\eeq 
Note that $p$ in Equation (\ref{eq:gamma_c}) is the pressure of the shell, and 
$p$ in Equation (\ref{eq:eta}) is the slope of the electron spectrum. 
When a fresh shell $\delta m$ is created at a shock front (FS or RS), 
$\gamma_c = +\infty$ is adopted as its initial value. 
We then solve Equation (\ref{eq:gamma_c}) numerically 
and find a subsequent evolution of $1/\gc$ (and hence, $\gc$) for the shell $\delta m$.

%--------------------------------------------------------

\subsection{Curvature effect and light curves} \label{section:curvature_effect}

\begin{figure}
\begin{center}
\includegraphics[width=10cm]{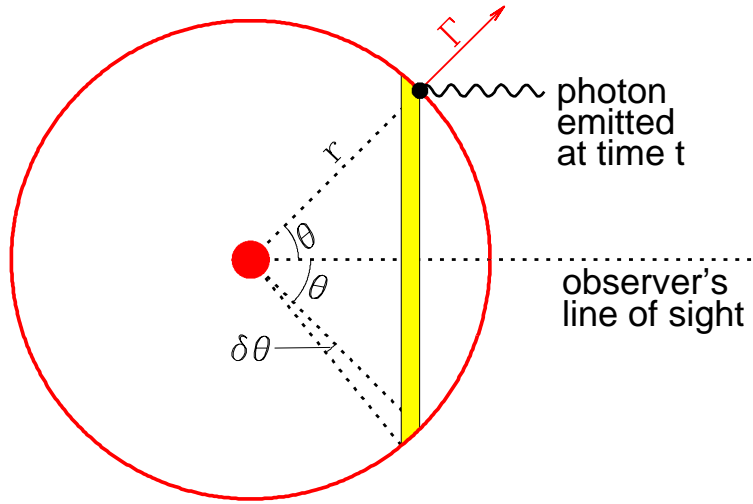}
\caption{
Schematic figure of a spherical shell at radius $r$ at time $t$ 
expanding with a Lorentz factor $\Gamma$. 
A photon emitted in the direction of the observer at time $t$ 
by an electron at a polar angle $\theta$ is received 
by the observer at an observer time $\tobs$ 
as given in Equation (\ref{eq:tobs}). 
A thin ring between $\theta$ and $\theta+\delta \theta$ is 
considered in order to derive an analytical expression for  
observed spectral flux $\delta F_{\nuobs}^{\rm obs}$; 
see Equation (\ref{eq:flux2}). 
} 
\label{fig:curvature}
\end{center}
\end{figure}

Consider a spherical shell $\delta m$ on the blast, 
which has radius $r$ at time $t$ 
expanding with a Lorentz factor $\Gb$. 
An observer is located in the positive $z$-direction 
at a large (cosmological) distance. 
See Figure~\ref{fig:curvature} for a schematic diagram.
An observer time $\tobs$ is set equal 
to zero when the observer detects the very first photon 
that was emitted at the explosion center at time $t=0$. 
Also, consider a thin ring on the shell $\delta m$, 
which is specified by a polar angle $\theta$ with respect to the $z$-axis. 
Then photons emitted from this ring at time $t$ will be detected 
by the observer at the observer time,
\be
\label{eq:tobs}
\tobs = \left[t-\frac{r}{c}\, \mu \right] (1+z),
\ee
where $\mu \equiv \cos \theta$ and $z$ is the cosmological redshift of the burst. 
The factor $(1+z)$ is introduced due to the expansion of the universe.

The spherical shell $\delta m$ contains a total of $\delta N$ electrons 
uniformly distributed over the shell, 
retaining a non-thermal spectrum of $\gm$, $\gc$, and slope $p$ 
(see Section~\ref{section:electron_spectrum}). 
Since the number of electrons on the thin ring 
(between $\theta$ and $\theta+\delta \theta$) is given 
as $(|\delta \mu|/2)\, \delta N$, 
the electron spectrum of the ring becomes
\be
\label{eq:ring_spec}
\tilde f(\gamma_e) = \frac{|\delta \mu|\, \delta N}{2}\, 
\frac{p-1}{\gm} \left[ \frac{\gamma_e}{\gm} \right]^{-p},
\ee
where we assumed $\gamma_e \gg 1$ and $\gm \gg 1$. 
Here the tilde on $\tilde f(\gamma_e)$ indicates the ring. 
Note that $\tilde f(\gamma_e)$ is subject to 
the thickness of the ring, $\delta \theta$ or $\delta \mu$.

We now focus on a single electron in the ring. On average, 
the electron is assumed to emit synchrotron photons isotropically\footnote{
See Beloborodov et al. (2011) for discussion of anisotropic emission in the fluid frame.
} 
at a single characteristic frequency $\nuflu$ in the fluid frame, 
\be
\label{eq:nuflu}
\nuflu = (0.15) \frac{q_e B}{m_e c}\, \gamma_e^2.
\ee
Here $q_e$ is the electric charge of the electron, 
and $B$ is the magnetic field of the shell. 
The synchrotron luminosity of the electron is given by 
\be
L_e = \frac{1}{6 \pi}\, \sigma_T c\, B^2\, \gamma_e^2.
\ee 
The photon frequency $\nulab$ in the lab. frame is 
different from $\nuflu$ due to radial expansion of the shell. 
Since the radial bulk motion of the electron has the angle $\theta$ with 
the observer's line of sight (see Figure~\ref{fig:curvature}),  
the frequency $\nulab$ in the direction of the observer is simply 
\be
\label{eq:nulab2}
\nulab(\theta) = \frac{\nuflu}{\Gb (1-\beta \cos \theta)}.
\ee 
The angular distribution of the energy of photons 
emitted by the electron in the direction of the observer 
is written in the lab. frame as  
\be
\left( \frac{d E_1}{d \Omega\, dt} \right)_{\rm lab} =
\frac{L_e}{4 \pi}\, \frac{1}{\Gb^4 (1-\beta \cos \theta)^3}.
\ee 
Here the subscript in $E_1$ refers to the {\it single} electron.

From Equations (\ref{eq:nuflu}) and (\ref{eq:nulab2}), 
we note that the Lorentz factors $\gm$ and $\gc$ 
of the electron spectrum correspond to the frequencies 
$\nu_m^{\rm lab}$ and $\nu_c^{\rm lab}$, respectively,
\beq
\label{eq:numlab}
\nu_m^{\rm lab} = (0.15)\, \frac{q_e B}{m_e c}\, \frac{\gm^2}{\Gb (1-\beta \mu)}, \\
\nu_c^{\rm lab} = (0.15)\, \frac{q_e B}{m_e c}\, \frac{\gc^2}{\Gb (1-\beta \mu)}.
\eeq
The electron with $\gamma_e=\gamma_{\nulab}$ has 
the frequency $\nulab$ in the direction of the observer, 
\be
\label{eq:nulab}
\nulab = (0.15)\, \frac{q_e B}{m_e c}\, \frac{\gamma_{\nulab}^2}{\Gb (1-\beta \mu)}.
\ee

The spherical shell emits photons continuously when it expands, 
but we think of a series of ``snapshots'' of the shell. 
We consider two consecutive snapshots (i.e., two calculation steps) 
separated by a time interval $\delta t$, and then assume that 
the shell accumulates its emission between two snapshots 
and emits all the accumulated energy instantaneously like a ``flash'' 
when it arrives at the second snapshot. 
In other words, the emission from the shell is viewed as 
a series of flashes.

When the shell flashes, the accumulated energy by 
the single electron during $\delta t$ is emitted. 
The energy emitted into a solid angle $\delta \Omega$ in the direction 
of the observer is given in the lab. frame as  
\be
\label{eq:deltaE1}
\delta E_1^{\rm lab} \equiv 
\left( \frac{d E_1}{d \Omega\, dt} \right)_{\rm lab}\, \delta \Omega\, \delta t = 
\frac{L_e}{4 \pi}\, \frac{\delta \Omega\, \delta t}{\Gb^4 (1-\beta \mu)^3}.
\ee 
The next step is then to calculate the emission from the entire thin ring; 
let $\delta \tilde E^{\rm lab}$ be the emission from the entire ring 
into the solid angle $\delta \Omega$ in the direction 
of the observer. 
With the definition of spectral energy 
$\delta \tilde E_{\nu}^{\rm lab} \equiv d(\delta \tilde E^{\rm lab})/d \nu$, 
we consider (e.g., B05) 
\beq
\nulab(\delta \tilde E_{\nulab}^{\rm lab}) 
&=& \left.\frac{d(\delta \tilde E^{\rm lab})}{d(\ln \nu)} \right|_{\nu=\nulab} 
 =  \left.\frac{d(\delta \tilde E^{\rm lab})}{2\, d(\ln \gamma_e)} 
    \right|_{\gamma_e=\gamma_{\nulab}} \nonumber \\
\label{eq:nu_deltaE}    
&=& \frac{1}{2}\, \gamma_{\nulab}\, \left. \frac{d(\delta \tilde E^{\rm lab})}{d\gamma_e} 
    \right|_{\gamma_e=\gamma_{\nulab}}.
\eeq
Here the second equality uses the relation 
$\delta (\ln \nulab) = 2\, \delta (\ln \gamma_{\nulab})$, 
which is verified from Equation~(\ref{eq:nulab}); 
for the instantaneous flash of the shell, 
a variation in $\nulab$ results only from $\delta \gamma_{\nulab}$ 
since the ring has a fixed $B$, $\Gb$, and $\mu$ instantaneously. 
Together with Equations (\ref{eq:ring_spec}) and (\ref{eq:deltaE1}), 
Equation~(\ref{eq:nu_deltaE}) yields 
\beq
\nulab(\delta \tilde E_{\nulab}^{\rm lab}) 
&=&  \frac{1}{2}\, \gamma_{\nulab}\, \left. \left[\delta E_1^{\rm lab}\; 
    \tilde f(\gamma_e) \right] \right|_{\gamma_e=\gamma_{\nulab}} \nonumber \\
\label{eq:nu_deltaE2}    
&=& \frac{1}{24\pi}\, \frac{p-1}{4\pi}\, \sigma_T c\, B^2 \gamma_{\nulab}^2\, 
    \frac{\delta \Omega\, \delta t\; |\delta \mu|\, \delta N}{\Gb^4(1-\beta \mu)^3} 
    \left[ \frac{\gamma_{\nulab}}{\gm} \right]^{1-p}.
\eeq 
We remark that Equation (\ref{eq:nu_deltaE2}) is valid only for 
$\nu_m^{\rm lab} < \nulab < \nu_c^{\rm lab}$ 
since the electron spectrum in Equation~(\ref{eq:ring_spec}) 
is valid only for $\gm < \gamma_{\nulab} < \gc$. 
Dividing Equation (\ref{eq:nu_deltaE2}) 
by $\nulab$ in Equation (\ref{eq:nulab}), 
we find the spectral energy emitted from the entire thin ring 
into the solid angle $\delta \Omega$ in the direction of the observer, 
\be
\label{eq:deltaE_nu}
\delta \tilde E_{\nulab}^{\rm lab} 
= \frac{5}{18\pi}\, \frac{p-1}{4\pi}\, \frac{\sigma_T\, m_e c^2\, B}{q_e}\, 
  \frac{\delta \Omega\, \delta t\; |\delta \mu|\, \delta N}{\Gb^3(1-\beta \mu)^2} 
  \left[ \frac{\gamma_{\nulab}}{\gm} \right]^{1-p}.
\ee
This energy is emitted instantaneously during the flash, 
but the thickness of the ring 
(between $\theta$ and $\theta+\delta \theta$) 
introduces a time interval $\delta \tilde \tlab$ 
along the observer's line of sight, 
\be
\label{eq:tlab}
\delta \tilde \tlab
=\frac{r}{c}\, \left[\cos \theta -\cos (\theta+\delta \theta) \right]
=\frac{r}{c}\, |\delta \mu|.
\ee
Hence, Equations (\ref{eq:deltaE_nu}) and (\ref{eq:tlab}) yield 
the spectral luminosity of the entire ring, 
\beq
\delta L_{\nulab}^{\rm lab}
&=& \frac{\delta \tilde E_{\nulab}^{\rm lab}}{\delta \tilde \tlab} \nonumber \\
\label{eq:deltaL_nu}
&=& \frac{5}{18\pi}\, \frac{p-1}{4\pi}\, \frac{\sigma_T\, m_e c^3\, B}{q_e\, r}\, 
    \frac{\delta \Omega\, \delta t\, \delta N}{\Gb^3(1-\beta \mu)^2} 
    \left[ \frac{\gamma_{\nulab}}{\gm} \right]^{1-p},
\eeq
which shines into the solid angle $\delta \Omega$ in the direction of the observer. 
Note that we do not include a tilde for $\delta L_{\nulab}^{\rm lab}$ 
since the thickness of the ring (namely, $\delta \mu$) cancels out here.

The burst is at a cosmological distance from the observer. 
For a flat $\Lambda$CDM universe,
the luminosity distance of a burst at redshift $z$ is given by 
\be
D_{\rm L} = \frac{c\,(1+z)}{H_0} 
\int_0^z \frac{dz^{\prime}}{\sqrt{\Omega_{\rm m} (1+z^{\prime})^3+\Omega_{\Lambda}}},
\ee
where $H_0=71$ km $\mbox{s}^{-1}$ $\mbox{Mpc}^{-1}$, 
$\Omega_{\rm m}=0.27$, and $\Omega_{\Lambda}=0.73$ (the concordance model). 
The photons are redshifted while traveling the cosmological distance, 
and the observed frequency $\nuobs$ is obtained by 
\be
\nuobs=\frac{\nulab}{1+z}.
\ee
The definition of the luminosity distance gives 
an observed spectral flux at $\nuobs$, 
\beq
\delta F_{\nuobs}^{\rm obs}
&=& \frac{(1+z)\,\delta L_{\nulab}^{\rm lab}}{D_{\rm L}^2\delta \Omega} \nonumber \\
\label{eq:flux}
&=& \frac{5(1+z)}{18\pi}\, \frac{p-1}{4\pi D_{\rm L}^2}\, \frac{\sigma_T\, m_e c^3\, B}{q_e\, r}\, 
    \frac{\delta N\, \delta t}{\Gb^3(1-\beta \mu)^2} 
    \left[ \frac{\nulab}{\nu_m^{\rm lab}} \right]^{(1-p)/2},
\eeq
where Equations (\ref{eq:numlab}) and (\ref{eq:nulab}) have been used 
to replace $\gamma_{\nulab}/\gm$ by $(\nulab/\nu_m^{\rm lab})^{1/2}$.

Finding $\mu=(c/r) \left[t-\tobs/(1+z) \right]$ 
from Equation (\ref{eq:tobs}) 
and substituting it into Equation (\ref{eq:flux}), 
we arrive at an analytical expression for 
$\delta F_{\nuobs}^{\rm obs}$ in terms of $\tobs$ and $\nuobs$,
\be
\label{eq:flux2}
\delta F_{\nuobs}^{\rm obs}(\tobs, \nuobs)
=\delta F_{\nuobs}^{\rm max} \times 
 \left[\frac{\nuobs}{\nu_m^{\rm obs}} \right]^{(1-p)/2},
\ee
where 
\beq
\delta F_{\nuobs}^{\rm max} &\equiv& 
\frac{5(1+z)}{18\pi}\,\frac{p-1}{4\pi D_{\rm L}^2}\,\frac{\sigma_T\, m_e c^3\,B}{q_e\, r}\, 
\frac{\delta N\; \delta t}
{\Gb^3 \left[1-\frac{c\beta}{r}\left(t-\frac{\tobs}{1+z}\right)\right]^2}, \\
\nu_m^{\rm obs} &=& 
\frac{0.15}{1+z}\, \frac{q_e B}{m_e c}\, 
\frac{\gm^2}{\Gb \left[1-\frac{c\beta}{r} \left(t-\frac{\tobs}{1+z}\right)\right]}. 
\eeq 
When the spherical shell of radius $r$ flashes at time $t$, 
the observer at cosmological distance receives 
its emission for a period of the observer time $\tobs$,
\be
\label{eq:tobs_period}
t-\frac{r}{c} \leq \frac{\tobs}{1+z} \leq t+\frac{r}{c}.
\ee
In other words, Equation (\ref{eq:flux2}) is to be evaluated 
only in this period of $\tobs$. For any $\tobs$ in this period, 
the observed spectral flux $\delta F_{\nuobs}^{\rm obs}$ 
at the observed frequency $\nuobs$ is expressed analytically. 
Thus we do not need to consider the ring any more.

Equation (\ref{eq:nu_deltaE2}) is valid for 
$\nu_m^{\rm lab} < \nulab < \nu_c^{\rm lab}$, 
and therefore Equation (\ref{eq:flux2}) is valid only for 
$\nu_m^{\rm obs} < \nuobs < \nu_c^{\rm obs}$, 
where 
\be
\label{eq:nuc_obs}
\nu_c^{\rm obs} = 
\frac{0.15}{1+z}\, \frac{q_e B}{m_e c}\, 
\frac{\gc^2}{\Gb \left[1-\frac{c\beta}{r} \left(t-\frac{\tobs}{1+z}\right)\right]}. 
\ee 
In general, the observed spectral flux at $\nuobs$ can be obtained as 
\beq
\label{eq:flux_general}
\delta F_{\nuobs}^{\rm obs}=
\delta F_{\nuobs}^{\rm max} \times
\left\{ 
\begin{array}{ll}
\left(\nuobs/\nu_m^{\rm obs} \right)^{1/3} 
\qquad & \mbox{for}\quad \nuobs < \nu_m^{\rm obs} < \nu_c^{\rm obs}, \\
\left(\nuobs/\nu_m^{\rm obs} \right)^{(1-p)/2} 
\qquad & \mbox{for}\quad \nu_m^{\rm obs} < \nuobs < \nu_c^{\rm obs}, \\
\left(\nuobs/\nu_c^{\rm obs} \right)^{1/3} 
\qquad & \mbox{for}\quad \nuobs < \nu_c^{\rm obs} < \nu_m^{\rm obs}, \\
\qquad 0
\qquad & \mbox{for}\quad \nu_c^{\rm obs} < \nuobs.
\end{array}
\right.
\eeq 
Equation (\ref{eq:flux_general}) needs to be summed over the shocked region 
as the blast is made of many different shells.

Here we recover the index $i$ to name our Lagrangian shells $\{\delta m^i\}$.
Each $\delta m^i$ is impulsively heated at some point by the FS or the RS 
and becomes a shocked shell on the blast. 
We denote these shocked shells by $\{\delta m_{\rm shd}^i\}$. 
As mentioned above, 
the shells $\{\delta m_{\rm shd}^i\}$ on the blast have their own individual 
radius $r^i$, number of electrons $\delta N^i$, magnetic field $B^i$, 
and the Lorentz factors $\gm^i$ and $\gc^i$.

Let us now use another index $j$ to specify the time $t^j$ 
of each calculation step (or flash). 
Solving for the blast wave dynamics at a calculation step with time $t^j$, 
we find the Lorentz factor $\Gamma(t^j)$ of the blast and 
the radii $\{r^i(t^j)\}$ of the shells $\{\delta m_{\rm shd}^i\}$. 
We also evaluate the shells' emission properties 
$\{B^i(t^j), \gm^i(t^j), \gc^i(t^j)\}$ at time $t^j$ 
(Sections~\ref{section:adiabatic_blast} and \ref{section:electron_spectrum}).

Then, Equation (\ref{eq:flux_general}) gives the spectral flux 
$\delta F_{\nuobs}^{i j\;\rm obs}$ 
of each shell $\delta m_{\rm shd}^i$ at time $t^j$. 
We sum it over all shocked shells $\{\delta m_{\rm shd}^i\}$ 
to find a flux 
$\delta F_{\nuobs}^{j\;\rm obs}
=\sum_{\{\delta m_{\rm shd}^i\}} \delta F_{\nuobs}^{i j\;\rm obs}$ 
of the entire blast at time $t^j$. 
Lastly, this needs to be summed over all flashes separated by time intervals 
$\{\delta t^j\}$ to give a total spectral flux,
\be
\label{eq:flux_sum}
F_{\nuobs}^{\rm obs}(\tobs, \nuobs)=
\sum\limits_{\{\delta t^j\}} 
\sum\limits_{\{\delta m_{\rm shd}^i\}} \delta F_{\nuobs}^{i j\;\rm obs} (\tobs, \nuobs). 
\ee
Note that two summations are not commutative 
since the FS and RS waves create new shocked shells as time goes. 
We find the ``FS emission'' (or the ``RS emission'') by taking the summation 
$\sum_{\{\delta m_{\rm shd}^i\}}$ over all shells only in the shocked ambient medium 
(or only in the shocked ejecta). 
For a fixed observed frequency $\nuobs$, Equation (\ref{eq:flux_sum}) gives 
light curves $F_{\nuobs}^{\rm obs}(\tobs)$ at $\nuobs$.
For a fixed observer time $\tobs$, Equation (\ref{eq:flux_sum}) yields 
flux spectra $F_{\nuobs}^{\rm obs}(\nuobs)$ at $\tobs$.

%--------------------------------------------------------
%
% Section 4  Numerical examples
%
%--------------------------------------------------------

\section{Numerical examples} \label{section:numerical_examples}

Now we present a total of 20 different numerical examples, which are named as follows; 
1a, 1b, 1c, 2a, 2b, 2c, 3a, 3b, 3c, 4a, 4b, 4c, 4d, 5a, 5b, 6a, 6b, 6c, 6d, and 6e. 
For all 20 examples, we keep the followings to be the same: 
(1) A constant density $\rho_1(r)/m_p=1~\mbox{cm}^{-3}$ is assumed for the ambient medium\footnote{
Although our method described in this paper allows us to study other types of environment, 
such as a stellar wind with $\rho_1 (r) \propto r^{-2}$, we will keep the same density 
$\rho_1(r)/m_p=1~\mbox{cm}^{-3}$ in all 20 examples. 
This is to focus on the study of the ejecta stratifications.
}. 
(2) The ejecta has a constant kinetic luminosity $\Lej(\tau)=L_0=10^{53}~\mbox{erg/s}$ for 
a duration of $\tau_b=10~\mbox{s}$, so that the total isotropic energy of the burst is to be 
$E_b=L_0\,\tau_b=10^{54}~\mbox{ergs}$.
(3) The burst is assumed to be located at a redshift $z=1$. 
(4) The emission parameters are 
$\epsilon_e=10^{-1}$, $\epsilon_B=10^{-2}$, and $p=2.3$ for the RS light curves, and 
$\epsilon_e=10^{-2}$, $\epsilon_B=10^{-4}$, and $p=2.3$ for the FS light curves.

Note that we have adopted different micophysics parameters for the FS and RS.
This is because the FS and RS shocked regions originate from different sources 
and the strengths of the two shocks can be significantly different. Indeed, 
afterglow modeling suggested that the RS can be more magnetized, 
and $\epsilon_e$ of the two shocks can also be different 
(e.g. Fan et al. 2002; Zhang et al. 2003; Kumar \& Panaitescu 2003).
Since the GRB central engine is likely magnetized, 
a natural consequence would be to invoke a larger $\epsilon_B$ in the RS. 
The bright optical flash observed in GRB 990123 (Akerlof et al. 1999) 
is likely related to such a case (Zhang et al. 2003).
Our emission parameters above are chosen such that the FS and RS fluxes 
are comparable to each other. Of course, either the FS or the RS emission 
could be further enhanced or suppressed by varying their parameters $\epsilon_B$ 
and/or $\epsilon_e$.

The only difference among examples then goes on the profile of the ejecta 
Lorentz factors $\Gej(\tau)$ as a function of ejection time $\tau$; i.e., the examples 
have a different ejecta stratification within the same duration $\tau_b=10~\mbox{s}$. 
Since we are mainly interested in the afterglow light curves, we ignore the initial 
variation of $\Gej$ and take a simple uniform profile of high Lorentz factors early on. 
Thus, for all 20 examples, we assume a constant Lorentz factor $\Gej=500$ for the initial 3 s. 
From 3 to 10 s, the examples have a decreasing profile of $\Gej$,\footnote{
It is natural to assume a decreasing profile of $\Gej$ 
since the internal shocks during the prompt emission phase tend to
smoothen the $\Gej$ distribution and lead to a decreasing $\Gej$ profile 
(otherwise, additional internal shocks would occur). 
}
exhibiting various types of the ejecta stratifications. 
Note that only a comparable amount, 70 \% of the burst energy, has been distributed 
over the shells with lower Lorentz factors in order to maintain a long-lived RS. 
Thus, the deceleration of the blast wave would deviate only mildly from the solution of BM76.

The examples with the same number in their names share a similar shape of the ejecta 
stratifications, and we categorize 20 examples into 7 different groups; 
(1a/1b/1c), (2a/2b/2c), (3a/3b/3c), (4a/4b/4c/4d), 
(1a/5a/5b), (6a/6b/4d), and (6c/6d/6e). 
This is to provide an efficient comparison among examples. 
Note that we use 1a and 4d twice in the comparisons.

\begin{figure}
\begin{center}
\includegraphics[width=10cm]{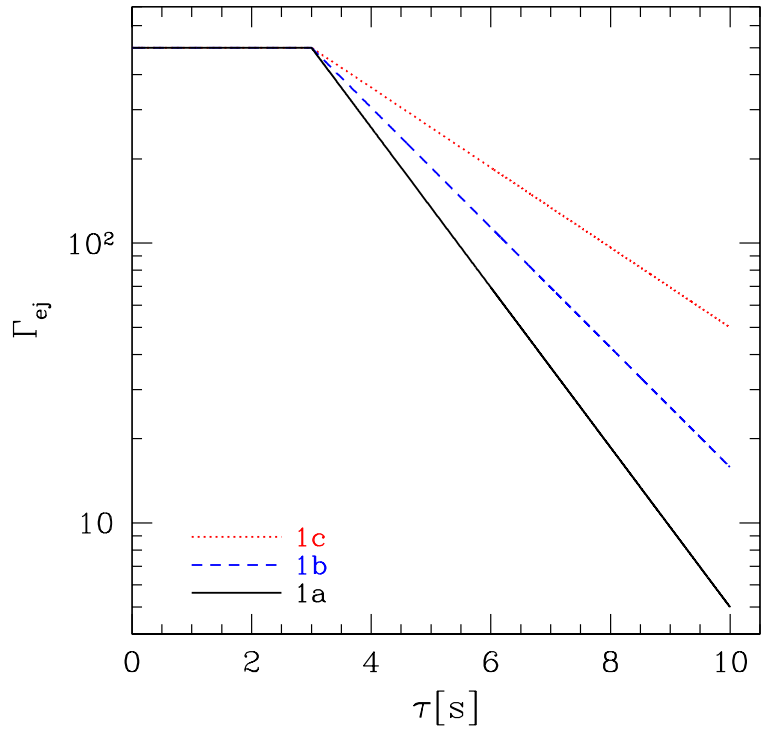}
\caption{
Ejecta stratifications for examples 1a, 1b, and 1c. 
The ejecta Lorentz factors $\Gej(\tau)$ are shown 
as a function of the ejection time $\tau$. 
} 
\label{fig:gej_1a1b1c}
\end{center}
\end{figure}

\begin{figure}
\begin{center}
\includegraphics[width=17cm]{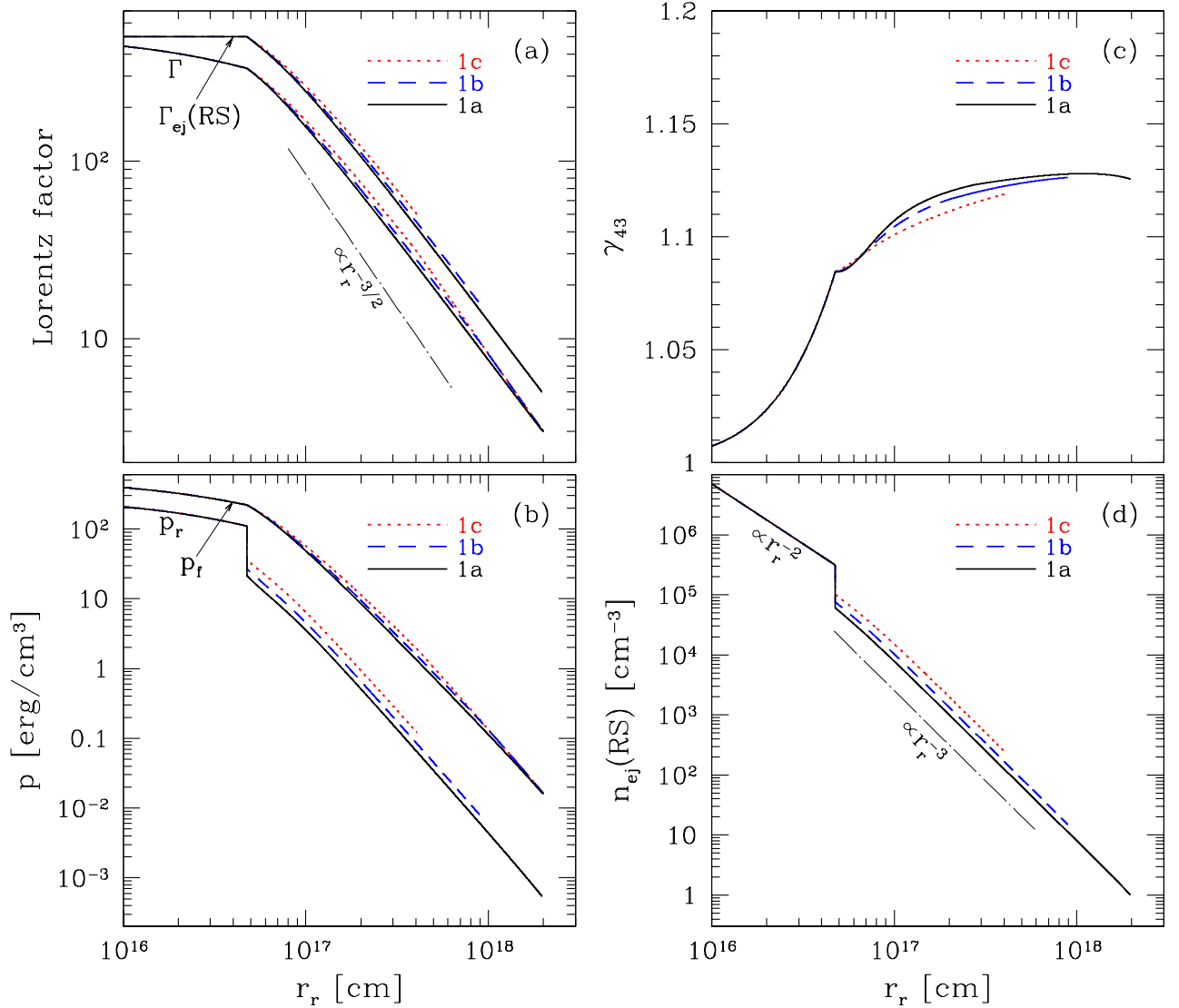}
\caption{
Blast wave dynamics for examples 1a, 1b, and 1c. 
In the panel (a), the curves denoted by $\Gamma$ show the Lorentz factor 
of the blast wave as a function of the radius $\rr$ of the RS, 
and the curves by $\GejRS$ show the Lorentz factor of the ejecta shell that gets 
shocked by the RS when the RS is located at radius $\rr$.
In the panel (b), the curves denoted by $\pf$ show the pressure at the FS, 
and the curves by $\pr$ show the pressure at the RS.
The panel (c) shows the relative Lorentz factor $\gamma_{43}$ across the RS wave, 
as is given by Equation (\ref{eq:g43}). The panel (d) shows the density 
$\nejRS = \rhoejRS / m_p$ of the ejecta shell, which enters the RS at radius $\rr$.
} 
\label{fig:dyn_1a1b1c}
\end{center}
\end{figure}

For each group of examples, we present 3 figures: 
(1) The 1st figure shows the ejecta stratifications (e.g., see Figure~\ref{fig:gej_1a1b1c}). 
The ejecta Lorentz factors $\Gej(\tau)$ are shown as a function of ejection time $\tau$, 
in different line (color) types. 
(2) The 2nd figure shows the blast wave dynamics of the examples (e.g., see Figure~\ref{fig:dyn_1a1b1c}). 
In the panel (a), the curves denoted by $\Gamma$ show the Lorentz factor 
of the blast wave as a function of the radius $\rr$ of the RS, 
and the curves by $\GejRS$ show the Lorentz factor of the ejecta shell that gets 
shocked by the RS when the RS is located at the radius $\rr$.
In the panel (b), the curves denoted by $\pf$ show the pressure at the FS 
as a function of the radius $\rr$, and the curves by $\pr$ show the pressure at the RS.
Since the ambient medium is assumed to have a constant density $\rho_1$, 
the $\pf$ curves resemble the $\Gamma$ curves; $\pf \propto \Gamma^2\, \rho_1$. 
The panel (c) shows the relative Lorentz factor $\gamma_{43}$ across the RS wave, 
as is given by Equation (\ref{eq:g43}). The panel (d) shows the density 
$\nejRS = \rhoejRS / m_p$ of the ejecta shell, which enters the RS wave at radius $\rr$.
We omit the panels (c) and (d) for the last two groups of examples.
(3) The 3rd figure shows the afterglow light curves (e.g., see Figure~\ref{fig:cRX_1a1b1c}). 
In the upper panel (a), we show the FS emissions in X-ray (1 keV) and {\it R} band 
as a function of the observer time $\tobs$. 
In the lower panel (b), we show the RS emissions in X-ray (1 keV) and {\it R} band.

Each group has a comparison point to help readers understand the effects of ejecta 
stratification on the FS and RS dynamics and further on the FS and RS afterglow light curves.

\subsection{Group (1a/1b/1c)} \label{section:1a1b1c}

Figure~\ref{fig:gej_1a1b1c} shows the ejecta stratifications of examples 1a, 1b, and 1c. 
For a duration from 3 to 10 s, 
the ejecta Lorentz factors $\Gej(\tau)$ decrease exponentially from 500 
to 5, $\sqrt{5 \times 50}$, and 50, respectively. 
Note that an exponential decrease implies 
$\frac{d}{d\tau}(\ln \Gej)=\Gej^{\prime}/\Gej=\mbox{const.}$

The blast wave dynamics of these three examples are shown in Figure~\ref{fig:dyn_1a1b1c}.
The $\GejRS$, $\pr$, $\gamma_{43}$, and $\rhoejRS$ curves vanish at an earlier time or radius 
for higher ending values of $\Gej$ (i.e., examples 1b and 1c), 
as the RS waves cross the end of ejecta. 
The $\Gamma$ curve of example 1a shows that its blast wave decelerates slightly slower than 
$\Gamma \propto \rr^{-3/2}$, the self-similar solution of BM76 (denoted by a dot-dashed line). 
The examples 1b and 1c have higher RS pressure $\pr$ than the example 1a, 
and therefore their $\Gamma$ curves deviate from BM76 even stronger than the case of 1a 
while their RS waves exist. However, once their RS waves vanish, 
their $\Gamma$ curves start to follow $\Gamma \propto \rr^{-3/2}$ 
as they should. As a result, all three $\Gamma$ curves completely agree with one another 
after all three RS waves disappear. 
This is expected because the same amount of burst energy $E_b$ has been injected 
into the blast waves which swept up the same amount of ambient medium 
out to a certain radius. 
In other words, three $\Gamma$ curves are shaped by three different plans or time schedules 
of ``spending'' the same energy budget $E_b$. Once the budget is used up, 
the outcome or the Lorentz factor of the blast waves should be the same. 
Also notice that three $\pf$ curves exhibit the same behaviors as $\Gamma$ curves, 
accordingly, since $\pf \propto \Gamma^2\, \rho_1$ and $\rho_1=\mbox{const}$.

Three $\GejRS$ curves stay close to the $\Gamma$ curves 
since the ejecta shells catch up with the blast waves only when $\Gej \sim \Gamma$; 
the resulting relative Lorentz factors $\gamma_{43}$ are shown in panel (c). 
A constant kinetic luminosity $\Lej(\tau)=L_0$ is assumed here, 
and thus Equation (\ref{eq:ejecta_density}) yields
$\rhoej(\tau,r) \propto r^{-2}$ when $\Gej^{\prime}(\tau)=0$.
Hence, $\rhoejRS = \rhoej(\tau_r, \rr) \propto \rr^{-2}$
while the RS waves sweep up the initial 3 s of ejecta shells 
with $\Gej=500$ (see panel(d)).
When the RS waves arrive at $\tau=3~\mbox{s}$, 
they encounter a discontinuity in the value of $\Gej^{\prime}(\tau)$, 
which results in a sudden drop in $\rhoej(\tau,r)$ 
across $\tau=3~\mbox{s}$; see Equation (\ref{eq:ejecta_density}).
Note that both $\rhoejRS$ and $\pr$ curves exhibit a sudden drop correspondingly.

When $\Gej^{\prime}(\tau)<0$, 
Equation (\ref{eq:ejecta_density}) simplifies 
\be
\label{eq:rhoej_simplified}
\rhoej(\tau, r) \propto \frac{\Lej(\tau)}{r^3}\, g_{\rm ej}(\tau)
\qquad
\mbox{if}
\quad
g_{\rm ej}(\tau) \ll \frac{r}{c\Gej^2},
\ee 
where we have defined
\be
g_{\rm ej}(\tau) \equiv \left[-\,\frac{d}{d\tau}(\ln \Gej) \right]^{-1}.
\ee 
For exponentially decreasing parts of $\Gej$ ($\tau > 3~\mbox{s}$), 
the examples 1a, 1b, and 1c have a constant value for $g_{\rm ej}(\tau)$, 
which equals 1.52, 2.03, and 3.04, respectively. 
Since $\frac{r}{c\Gej^2} \sim \tobs$, 
Equation (\ref{eq:rhoej_simplified}) is applicable here, 
and yields $\rhoej(\tau, r) \propto r^{-3}$. 
Hence, $\rhoejRS = \rhoej(\tau_r, \rr) \propto \rr^{-3}$
while the RS waves sweep up these exponential parts of ejecta shells 
(see panel (d))\footnote{
It is evident here that a radial spread-out of a stratified ejecta 
induces a lower density ejecta-flow 
than a non-stratified ejecta with $\Gej=\mbox{const.}$
due to its gradient $\Gej^{\prime}(\tau)$. }.
Then, the $\pr$ curves roughly follow $\pr \propto \rr^{-3}$, 
as the $\gamma_{43}$ curves mildly increase at around the value 1.1. 
Note that the $\pf$ curves also roughly follow $\pf \propto \rr^{-3}$, 
since the $\Gamma$ curves deviate only mildly from $\Gamma \propto \rr^{-3/2}$.

Therefore, it is not surprising that both the FS and RS light curves exhibit 
power-law declines with a very similar value of temporal indices (see Figure~\ref{fig:cRX_1a1b1c}).
Note that our choice of emission parameters $\epsilon_e$ and $\epsilon_B$ produces 
the RS afterglow emissions at a comparable flux level to the FS emissions. 
An abrupt decline of the RS light curves after a rising phase is due to 
a sudden drop in the $\pr$ curves (mentioned above); 
this will become clear below with the group of examples 1a, 5a, and 5b.

Consequences of spending the energy budget in three different ways 
can be evidently seen in both the FS and RS light curves.
In particular, the RS waves of examples 1b and 1c vanish at an earlier time or radius, 
and thus their RS light curves exhibit a temporal break and steepen afterwards.
Even after the RS waves disappear, the RS light curves still get contributions 
from previously shocked Lagrangian shells whose $\nu_c^{\rm obs}$ (Equation (\ref{eq:nuc_obs})) 
is still higher than $\nuobs$; namely, the RS light curves are produced by 
residual emissions plus high latitude emissions.
In the case of examples 1b and 1c, the RS light curves are essentially 
governed by high latitude emissions after the temporal break. 
Notice that this RS break looks like a ``jet break'', which 
has been interpreted to be caused by a collimation of ejected outflow.

\subsection{Group (2a/2b/2c)}

\begin{figure}
\begin{center}
\includegraphics[width=10cm]{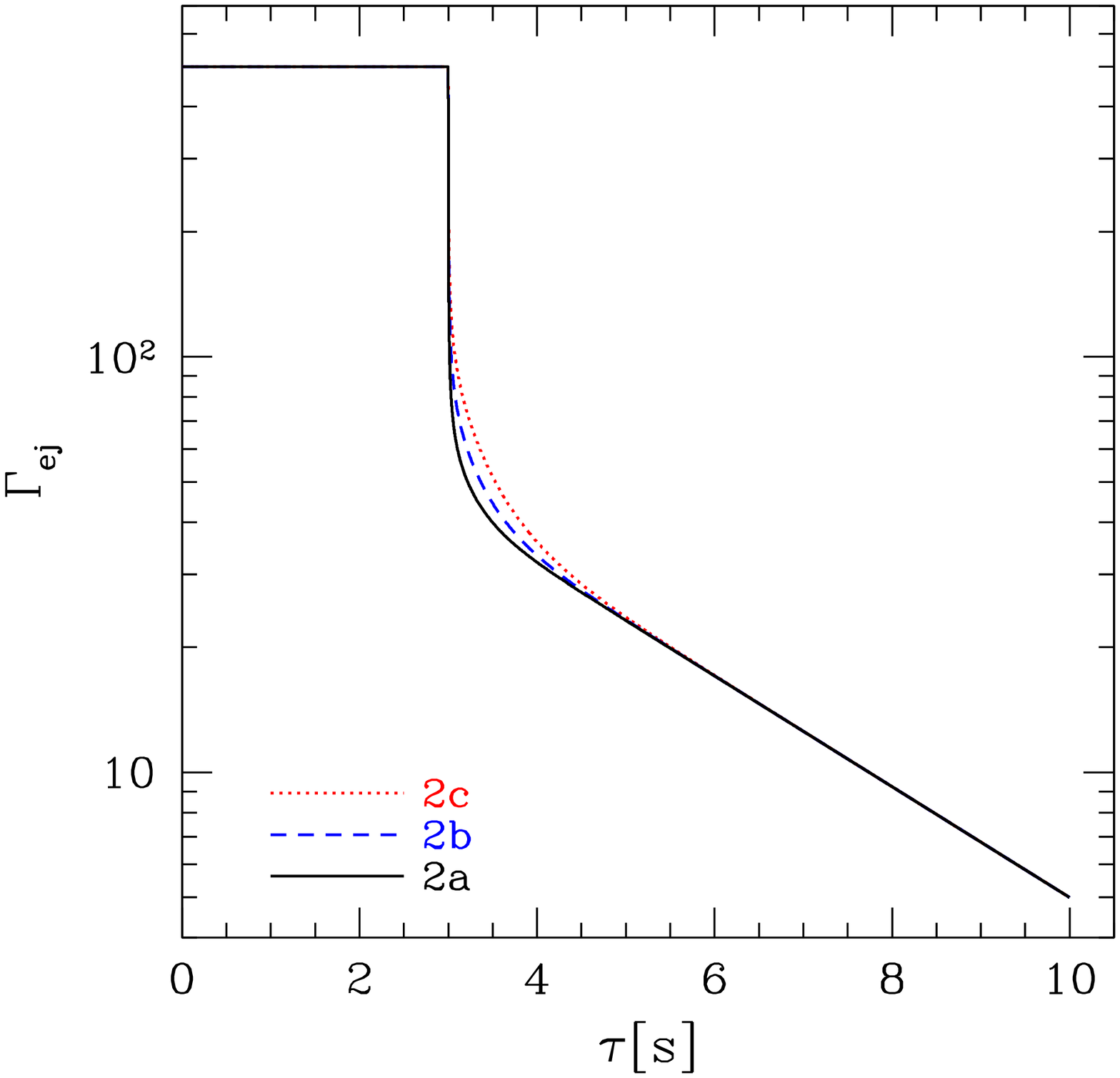}
\caption{
Same as in Figure~\ref{fig:gej_1a1b1c}, but for examples 2a, 2b, and 2c.
}
\label{fig:gej_2a2b2c} 
\end{center}
\end{figure}

The ejecta stratifications of examples 2a, 2b, and 2c are shown in Figure~\ref{fig:gej_2a2b2c}. 
After a steep decrease at $\tau=3~\mbox{s}$, the $\Gej$ profiles transition to an exponential decrease, 
displaying three different concave shapes.

\begin{figure}
\begin{center}
\includegraphics[width=17cm]{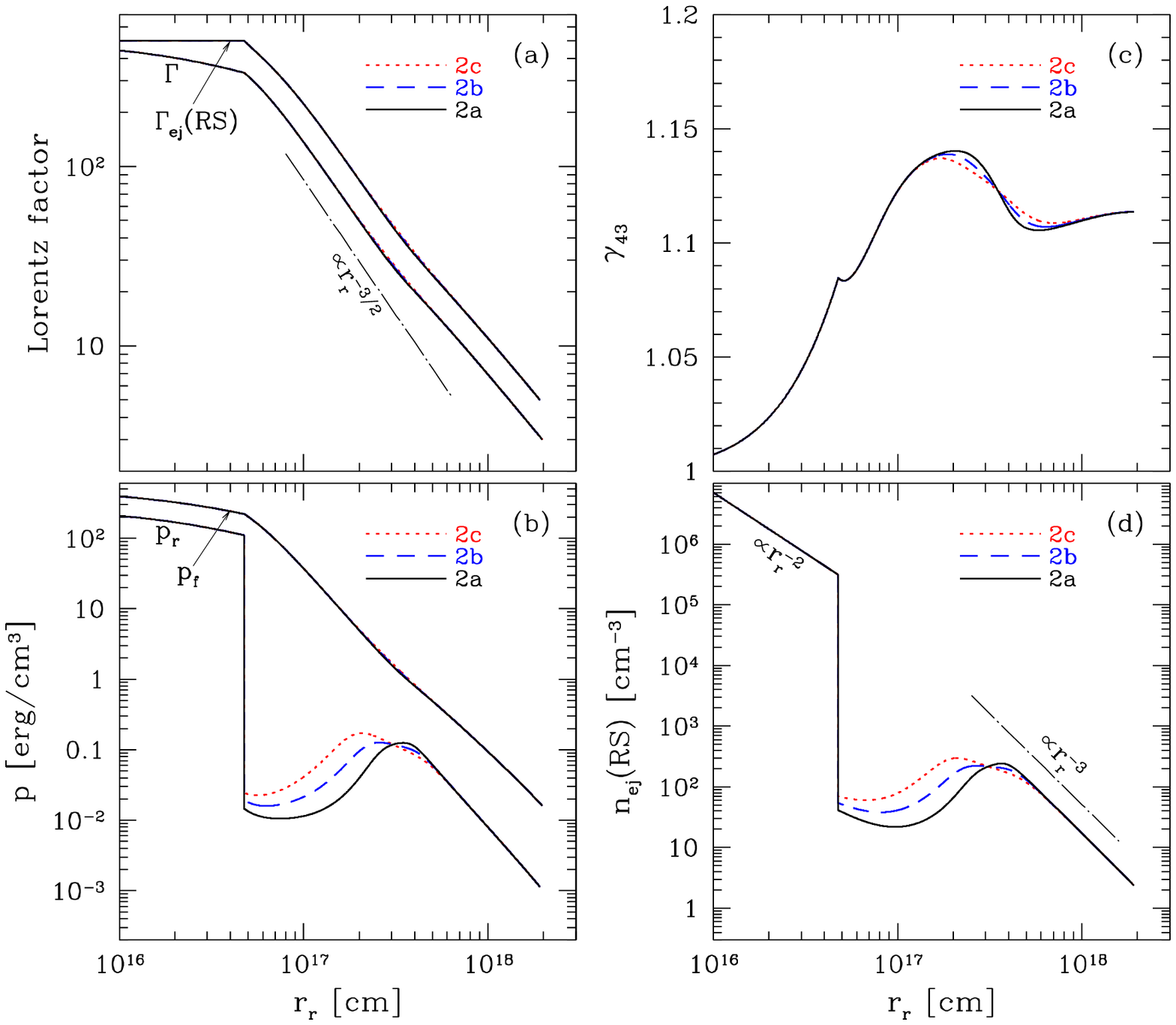}
\caption{
Same as in Figure~\ref{fig:dyn_1a1b1c}, but for examples 2a, 2b, and 2c.
}
\label{fig:dyn_2a2b2c} 
\end{center}
\end{figure}

A large gradient $\Gej^{\prime}(\tau)$ at $\tau=3+\varepsilon$ ($\varepsilon \ll 1$) 
results in a very small value for $g_{\rm ej}(\tau)$. 
The $g_{\rm ej}(\tau)$ then gradually increases as it goes through the concave shapes, 
and becomes a constant value of 3.26 when it continues on the exponential parts. 
Thus, Equation (\ref{eq:rhoej_simplified}) is applicable here. 
Hence, $\rhoejRS = \rhoej(\tau_r,\rr) \propto \rr^{-3}\, g_{\rm ej}(\tau_r)$.
Note that, when $g_{\rm ej}(\tau_r)$ rises, it competes with a decaying term $\rr^{-3}$ 
to determine the density profile $\rhoejRS$ of ejecta shells entering the RS waves. 
As shown in Figure~\ref{fig:dyn_2a2b2c}, the $\rhoejRS$ curves exhibit a rising phase 
and then eventually follow $\rhoejRS \propto \rr^{-3}$ when the RS waves sweep up 
the exponential parts of ejecta shells. 
A local maximum of the $\rhoejRS$ curves is located at different radii $\rr$ 
as a result of the competition between $g_{\rm ej}(\tau_r)$ and $\rr^{-3}$.

The $\pr$ curves resemble the $\rhoejRS$ curves; 
a significant drop and then a gradual recovery, 
followed by a power-law decline, $\pr \propto \rr^{-3}$ roughly 
(see panel (b)). 
While the RS pressures $\pr$ are weakened temporarily, 
the blast waves satisfy $\pr \ll \pf$, 
and therefore the $\Gamma$ curves follow $\Gamma \propto \rr^{-3/2}$. 
When the $\pr$ curves start to recover from the drop, 
the $\Gamma$ curves start to deviate from the self-similar solution.
Note that three $\Gamma$ curves show a very minor difference 
because a relatively large difference in $\pr$ curves during the recovery phase 
does not have much meaning to the $\Gamma$ curves since $\pr \ll \pf$. 
Three $\pf$ curves also show a negligible difference accordingly.

Hence, the FS light curves do not exhibit any noticeable difference 
(see Figure~\ref{fig:cRX_2a2b2c}). 
On the other hand, three different $\pr$ curves, of course, show up 
in the RS light curves.
During the temporary weakening of the RS waves, 
their X-ray light curves are dominated by high latitude emissions 
since the synchrotron frequencies $\nu_c$ corresponding to the
cutoff Lorentz factors $\gc$ of previously shocked Lagrangian shells 
are already below 1 keV. 
However, the frequencies $\nu_c$ stay above the {\it R} band for some time 
for the given values of emission parameters; recall that the Lorentz factor $\gc$ 
of each Lagrangian shell is determined by Equation (\ref{eq:gamma_c}), 
which involves $\epsilon_e$, $\epsilon_B$, $p$, etc. 
Therefore, even when the RS waves have been weakened significantly, 
the residual emissions from these previously shocked shells 
still contribute to the {\it R} band, and produce the {\it R} band light curves 
with a shallower decline (temporal index $\alpha \sim 1$) 
than the X-ray light curves of high latitude emissions ($\alpha \sim 3$)\footnote{
Note that the high latitude emission from an end of the prompt internal shocks 
would also naturally produce a similar steep decline in X-ray (Hasco\"et et al. 2012).
}.

During the recovery phase of the RS waves, 
their increasing $\pr$ curves are combined with decreasing $\Gamma$ curves, 
and as a consequence, the X-ray light curves exhibit a flattening phase (``plateau''). 
Note that the detailed shape of this plateau phase is inherited from the $\pr$ curves 
that are shaped by the concave parts of $\Gej$ profiles (Figure~\ref{fig:gej_2a2b2c}). 
This recovery of the RS waves is not fully shown up in the {\it R} band light curves 
since it is buried under the residual emissions (mentioned above). 
For a smaller value of $\epsilon_B$ in the RS, the electrons in the shocked shells 
would cool more slowly, and their resulting residual emissions would continue 
for a longer period of time in the {\it R} band, 
thus enhancing the chromaticity between the X-ray and {\it R} band.

As the RS waves continue on the exponential parts of $\Gej$ profiles, 
the $\pr$ curves roughly follow $\pr \propto \rr^{-3}$, 
and the RS light curves show the usual power-law decline with a temporal 
index ($\alpha \sim 1$), very similar to that of the FS light curves.
A temporal break at the end of the X-ray plateau phase marks the end of 
a rising phase of the $\pr$ curves, and therefore this ``dynamically-caused'' 
temporal break does not involve a spectral evolution across the break.

\subsection{Group (3a/3b/3c)}

\begin{figure}
\begin{center}
\includegraphics[width=10cm]{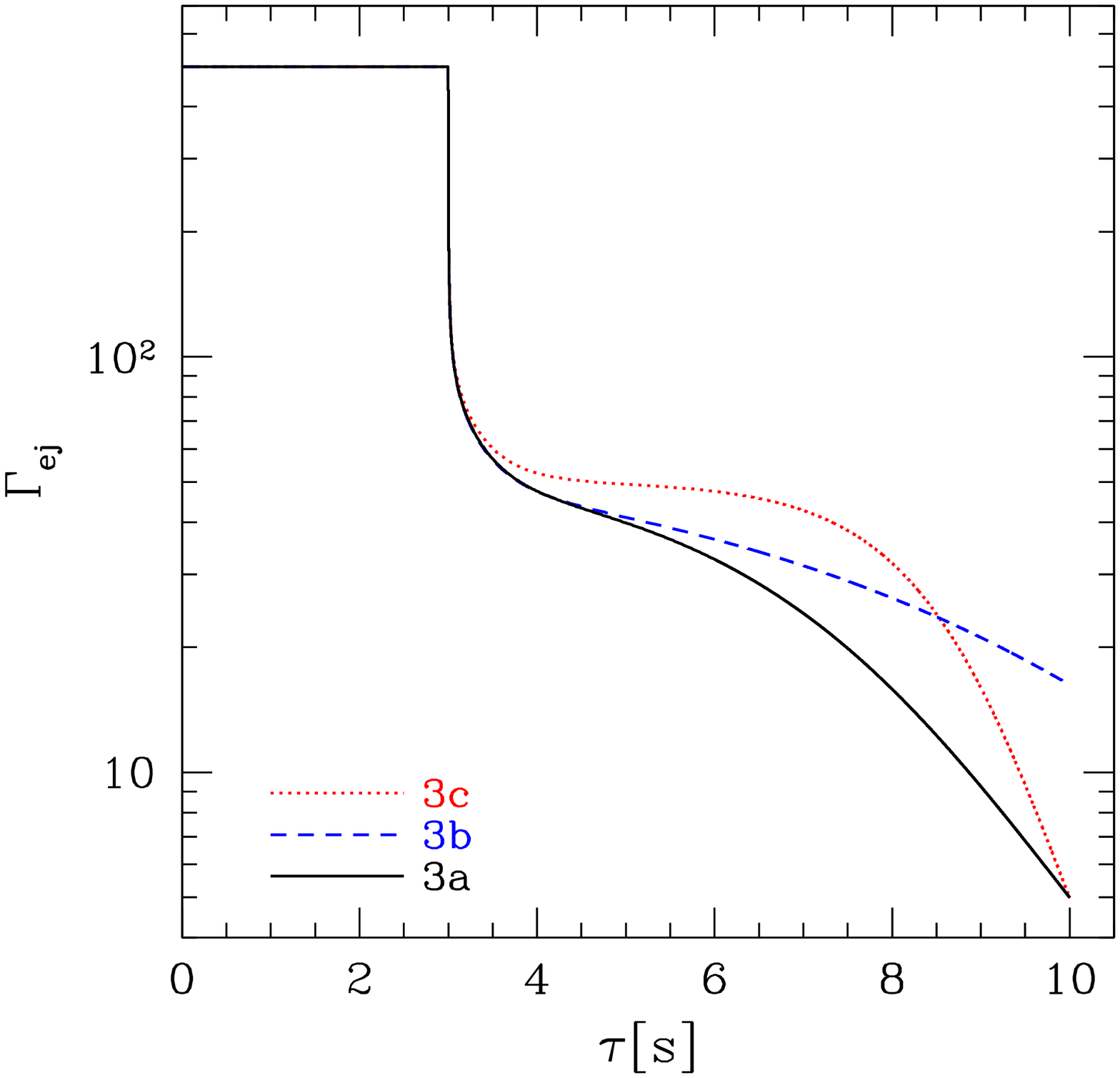}
\caption{
Same as in Figure~\ref{fig:gej_1a1b1c}, but for examples 3a, 3b, and 3c.
}
\label{fig:gej_3a3b3c} 
\end{center}
\end{figure}

The $\Gej$ profiles of examples 3a, 3b, and 3c are composed of 
a steep drop and a concave part, followed by a convex part (see Figure~\ref{fig:gej_3a3b3c}). 
The example 3b has a higher ending value of $\Gej$.

\begin{figure}
\begin{center}
\includegraphics[width=17cm]{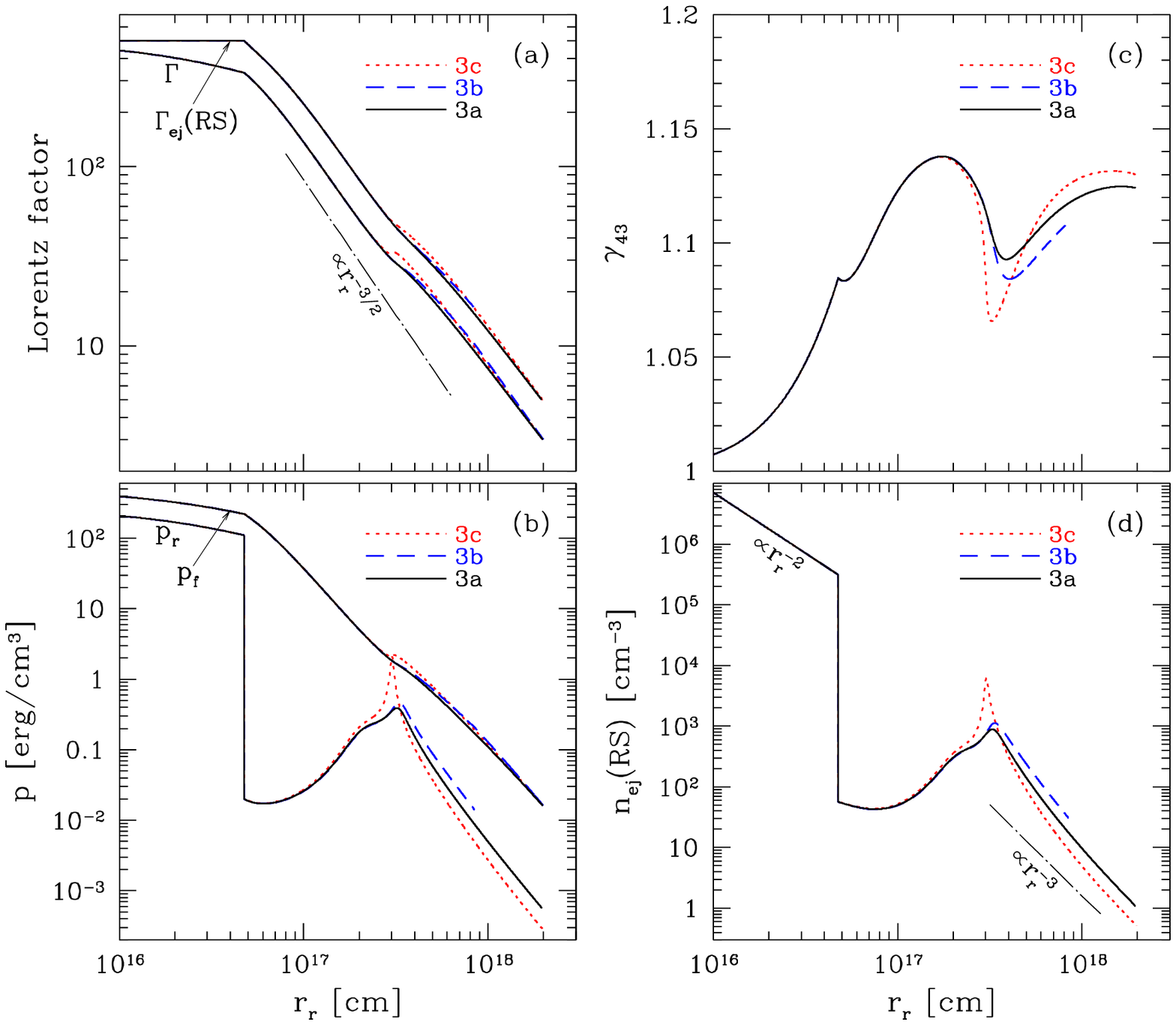}
\caption{
Same as in Figure~\ref{fig:dyn_1a1b1c}, but for examples 3a, 3b, and 3c.
}
\label{fig:dyn_3a3b3c} 
\end{center}
\end{figure}

The $g_{\rm ej}(\tau_r)$ functions increase until the end of concave parts 
(i.e., inflection points), and then decrease afterwards through convex parts.
Thus, the $\rhoejRS$ curves rise, exhibit a local maximum, 
and then decay afterwards (see Figure~\ref{fig:dyn_3a3b3c}).
Since  $\rhoejRS \propto \rr^{-3}\, g_{\rm ej}(\tau_r)$, 
the location of this local maximum does not exactly agree with, 
but roughly coincides with, the inflection point of $\ln \Gej$. 
Since the functions $g_{\rm ej}(\tau_r)$ decrease after the inflection point, 
the $\rhoejRS$ curves exhibit a more pronounced local maximum and decay faster afterwards 
than the case of examples 2a, 2b, and 2c.

In particular, the example 3c shows a well-pronounced local maximum 
in the $\rhoejRS$ and $\pr$ curves (Figure~\ref{fig:dyn_3a3b3c}), 
which appears in the RS light curves as a small flaring activity 
at the end of the plateau phase (Figure~\ref{fig:cRX_3a3b3c}). 
The examples 3a and 3b also show a clear temporal break at the end of the X-ray plateau phase. 
As the RS wave of example 3b vanishes at an earlier time or radius, 
its RS light curves display an additional temporal break at $\tobs \sim 10^5~\mbox{s}$.

\subsection{Group (4a/4b/4c/4d)}

\begin{figure}
\begin{center}
\includegraphics[width=10cm]{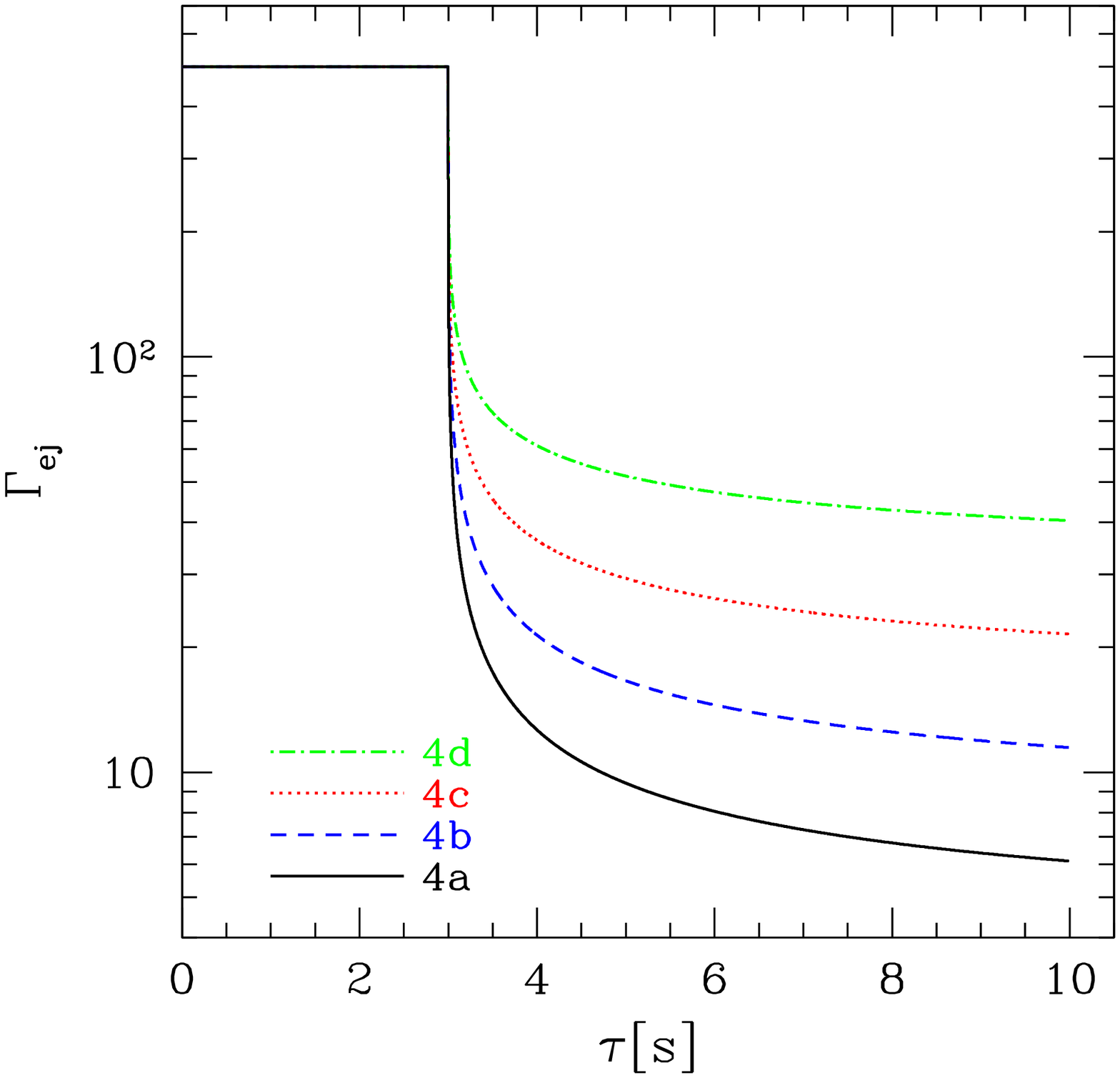}
\caption{
Same as in Figure~\ref{fig:gej_1a1b1c}, but for examples 4a, 4b, 4c, and 4d.
}
\label{fig:gej_4a4b4c4d} 
\end{center}
\end{figure}

The ejecta stratifications of examples 4a, 4b, 4c, and 4d have the same concave shape 
with various ending values that are equally spaced (Figure~\ref{fig:gej_4a4b4c4d}).

\begin{figure}
\begin{center}
\includegraphics[width=17cm]{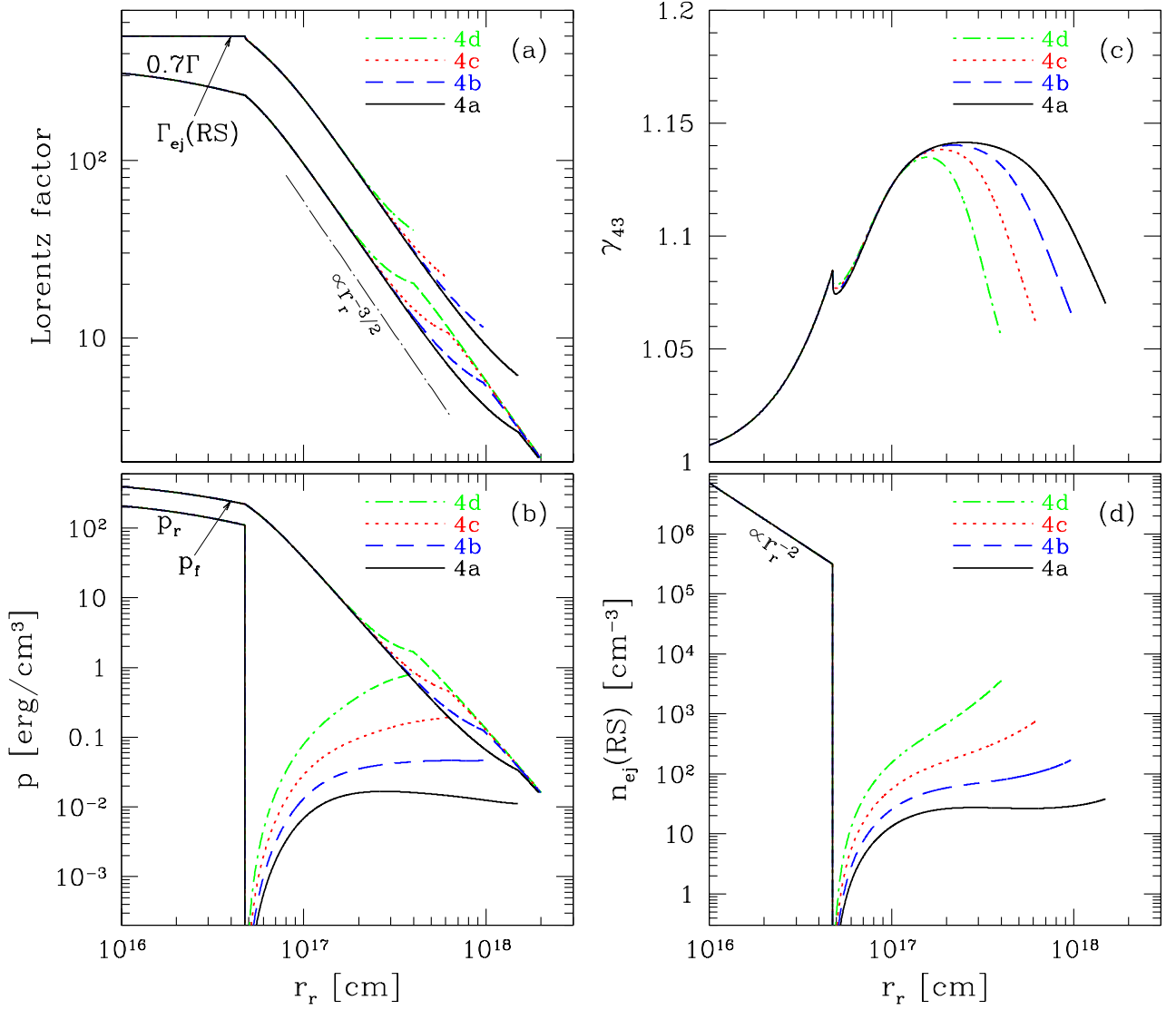}
\caption{
Same as in Figure~\ref{fig:dyn_1a1b1c}, but for examples 4a, 4b, 4c, and 4d. 
The $\Gamma$ curves are multiplied by a factor of 0.7, 
in order to avoid an overlap with the $\GejRS$ curves.
}
\label{fig:dyn_4a4b4c4d} 
\end{center}
\end{figure}

Initially, the $\Gamma$ curves follow $\Gamma \propto \rr^{-3/2}$ when $\pr \ll \pf$, 
and then start to deviate from it as the $\pr$ curves recover 
from a weakening (Figure~\ref{fig:dyn_4a4b4c4d}). 
Soon afterwards, the RS waves vanish, and their $\Gamma$ curves again 
follow $\Gamma \propto \rr^{-3/2}$. 
All four $\Gamma$ curves, of course, agree with one another 
after the RS waves disappear.

The $\pf$ curves resemble the $\Gamma$ curves, 
and the FS light curves show corresponding behaviors (Figure~\ref{fig:cRX_4a4b4c4d}).
During the recovery phase of the $\pr$ curves, 
their X-ray light curves produce a plateau phase. 
This plateau phase ends as the RS waves cross the end of ejecta. 
Therefore, the X-ray light curves exhibit a sudden steep decline ($\alpha \sim 3$) 
beyond this plateau phase. 
Note that, the less shallow the plateau phase is, the longer it is.
While the example 4d has an almost flat plateau phase ($\alpha \sim 0$), 
the example 4a has a mild plateau ($\alpha \sim 0.5$), 
which is so long that its end can not be seen here until $\tobs=10^6~\mbox{s}$.

\subsection{Group (1a/5a/5b)} \label{section:1a5a5b}

\begin{figure}
\begin{center}
\includegraphics[width=10cm]{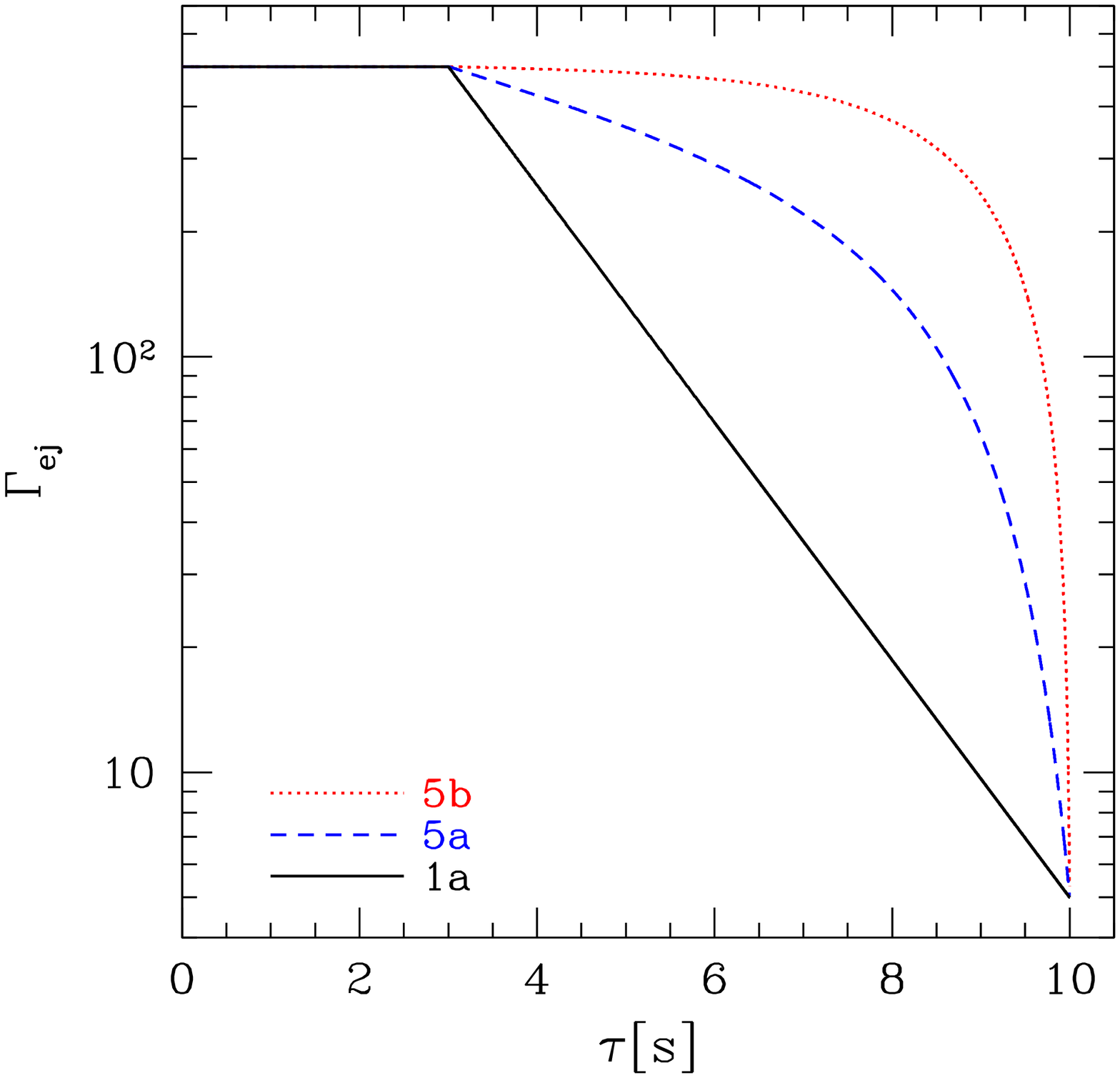}
\caption{
Same as in Figure~\ref{fig:gej_1a1b1c}, but for examples 1a, 5a, and 5b.
} 
\label{fig:gej_1a5a5b} 
\end{center}
\end{figure}

New examples 5a and 5b are compared to a previous example 1a here (Figure~\ref{fig:gej_1a5a5b}). 
The examples 5a and 5b show no steep drop at $\tau=3~\mbox{s}$. 
Instead, their $\Gej$ profiles are made of a single convex distribution.

\begin{figure}
\begin{center}
\includegraphics[width=17cm]{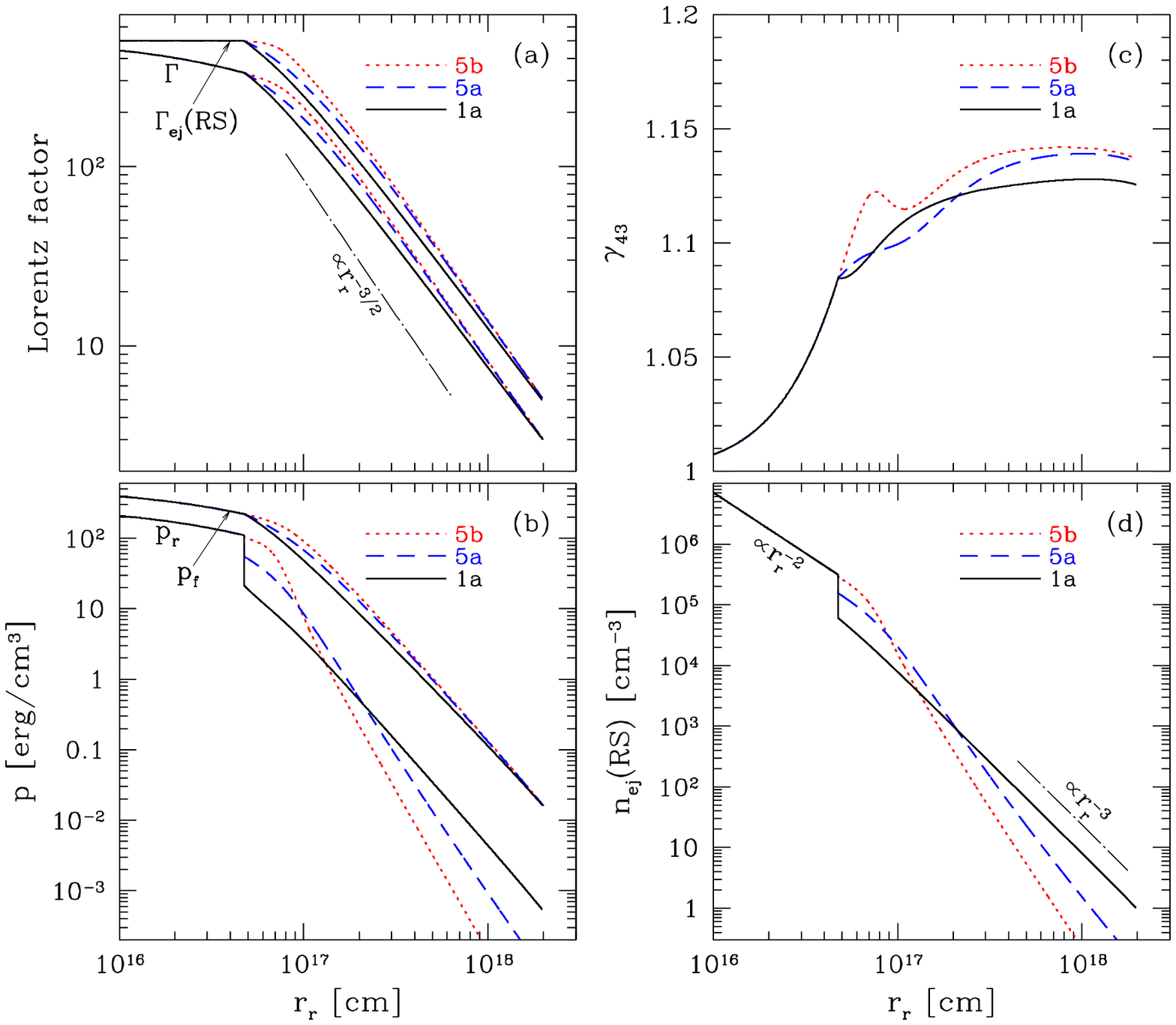}
\caption{
Same as in Figure~\ref{fig:dyn_1a1b1c}, but for examples 1a, 5a, and 5b.
} 
\label{fig:dyn_1a5a5b} 
\end{center}
\end{figure}

The example 5b has no discontinuity in $\Gej^{\prime}(\tau)$ across $\tau=3~\mbox{s}$, 
and therefore, its $\rhoejRS$ curve shows no sudden drop 
at $\tau=3~\mbox{s}$ (Figure~\ref{fig:dyn_1a5a5b}). 
When $\Gej^{\prime}(\tau) \sim 0$ beyond $\tau=3~\mbox{s}$, 
Equation (\ref{eq:ejecta_density}) roughly yields $\rhoej(\tau,r) \propto r^{-2}$, 
and its simplified version, Equation (\ref{eq:rhoej_simplified}), is not valid; 
hence, $\rhoejRS = \rhoej(\tau_r,\rr) \propto \rr^{-2}$. 
For the convex part of $\Gej$ profile, 
the $g_{\rm ej}(\tau)$ function keeps decreasing as $\tau$ increases, 
and as a result, Equation (\ref{eq:rhoej_simplified}) becomes applicable afterwards; 
$\rhoejRS \propto \rr^{-3}\, g_{\rm ej}(\tau_r)$. 
Thus, the $\rhoejRS$ curve starts to decrease faster than that of example 1a, 
for which $\rhoejRS \propto \rr^{-3}$.

The $\pr$ curves resemble the $\rhoejRS$ curves (panel (b)).
Since the $\pr$ curve of example 5a is initially higher than that of example 1a, 
its blast wave deceleration (i.e., $\Gamma$ curve) is delayed 
more significantly than that of example 1a. 
The $\Gamma$ curve of example 5b follows $\Gamma \propto \rr^{-3/2}$ afterwards 
when $\pr \ll \pf$.

The $\pf$ curves and the FS light curves show the same behaviors 
correspondingly (Figure~\ref{fig:cRX_1a5a5b}). 
The RS light curves also exhibit a similar behavior, 
but with a stronger variation in their temporal indices, 
since the $\pr$ curves have stronger variation than the $\pf$ curves. 
Note that the example 5b has no abrupt decline in its RS light curves 
after a rising phase since it has no sudden drop in its $\pr$ curve. 
This group of examples demonstrates that the FS and RS light curves 
can exhibit various temporal indices without varying the slope $p$ of 
electron spectrum. 
In particular, this becomes significant at late times in the RS light curves.

\subsection{Group (6a/6b/4d)}

\begin{figure}
\begin{center}
\includegraphics[width=10cm]{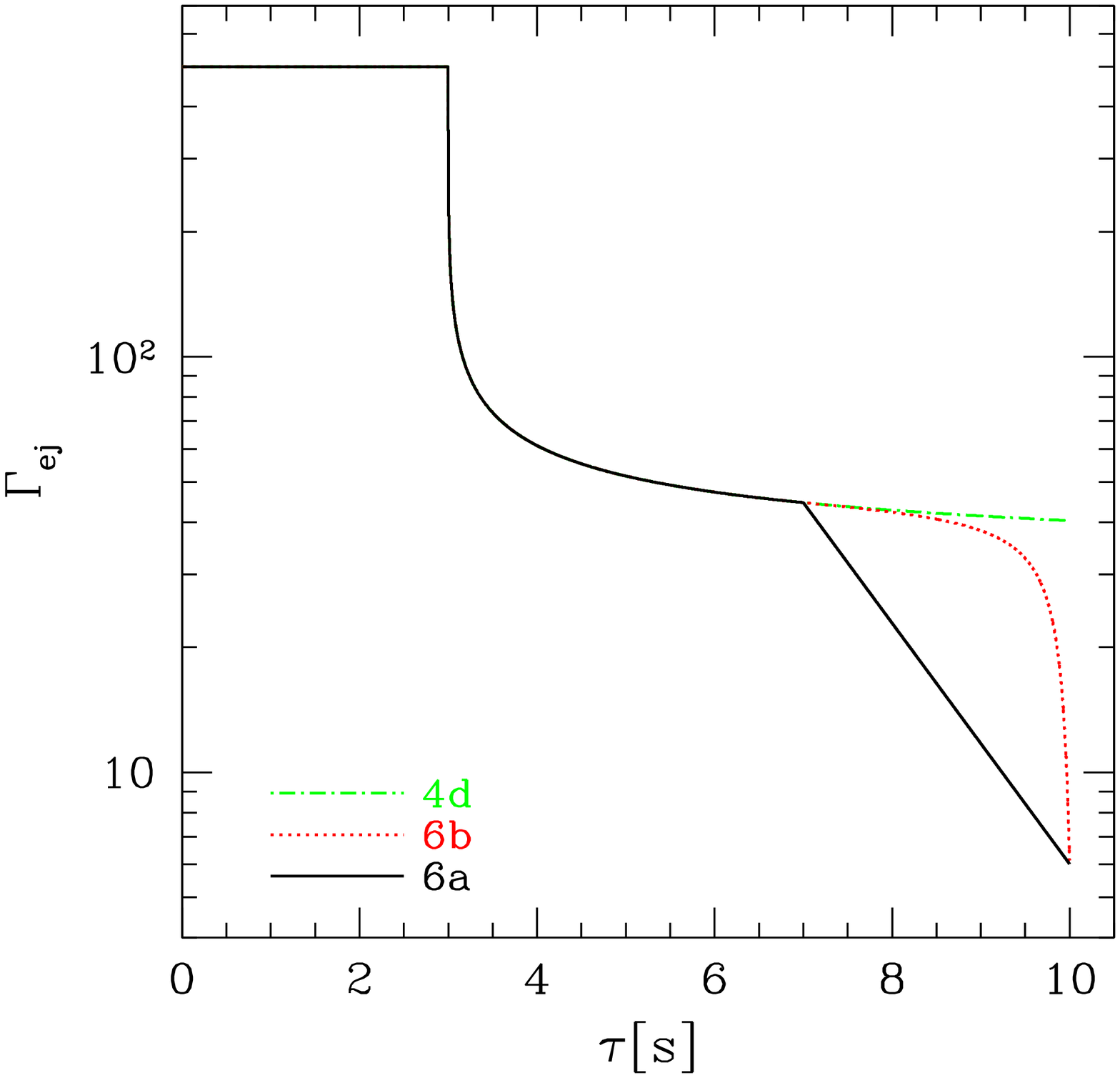}
\caption{
Same as in Figure~\ref{fig:gej_1a1b1c}, but for examples 6a, 6b, and 4d.
} 
\label{fig:gej_6a6b4d} 
\end{center}
\end{figure}

New examples 6a and 6b are compared to a previous example 4d (Figure~\ref{fig:gej_6a6b4d}).
Three examples have the same $\Gej$ profile up until $\tau=7~\mbox{s}$. 
From $\tau=7~\mbox{s}$, the example 6a has an exponential decrease 
while the example 6b continues with a convex profile.

\begin{figure}
\begin{center}
\includegraphics[width=10cm]{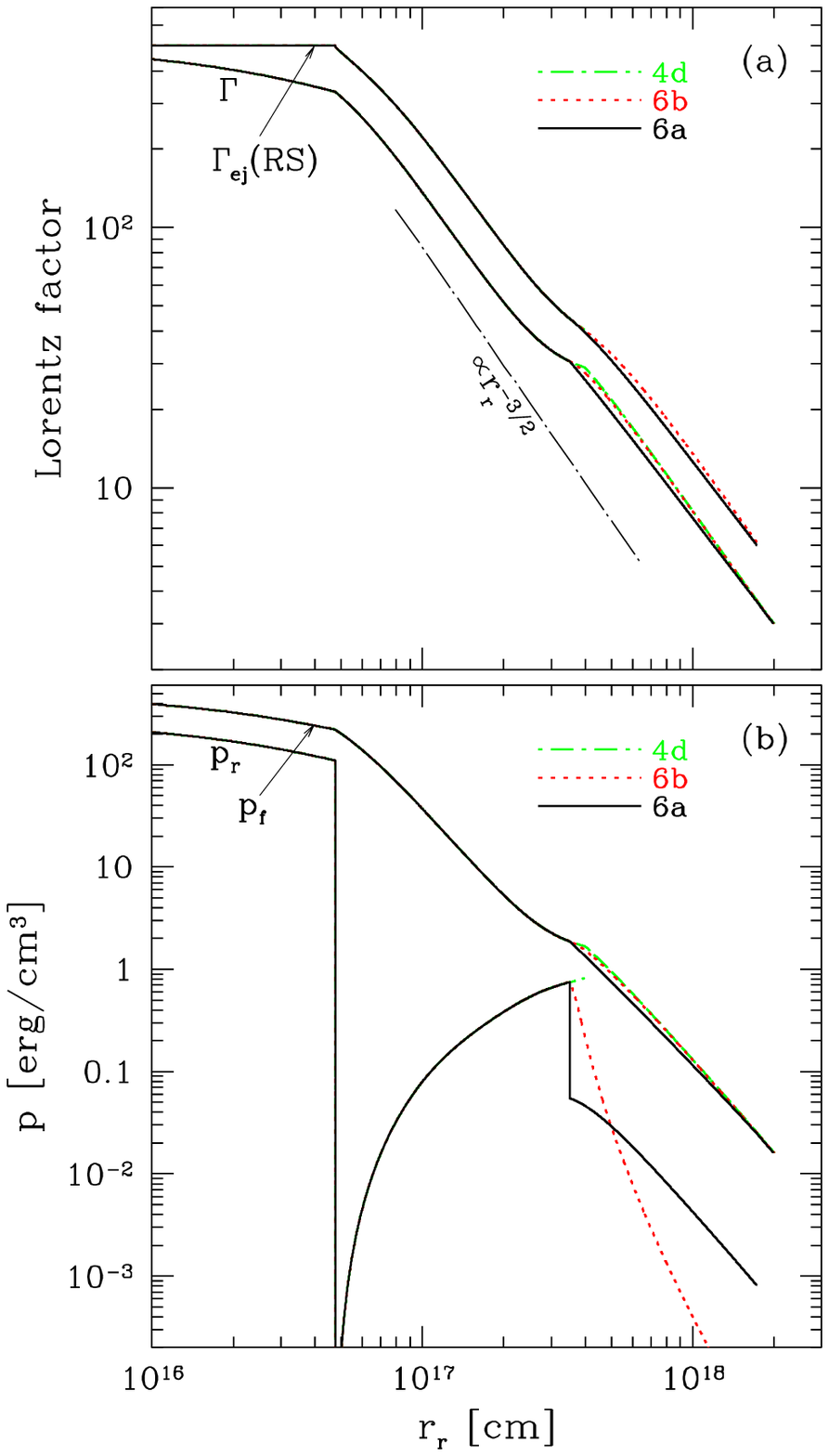}
\caption{
Same as in Figure~\ref{fig:dyn_1a1b1c}, but for examples 6a, 6b, and 4d.
} 
\label{fig:dyn_6a6b4d} 
\end{center}
\end{figure}

The $\pr$ curve of example 6a has another sudden drop at $\tau=7~\mbox{s}$, 
since $\Gej^{\prime}(\tau)$ is not continuous there (Figure~\ref{fig:dyn_6a6b4d}). 
The $\pr$ curve then roughly follows $\pr \propto \rr^{-3}$ 
as the RS wave sweeps up the exponential part of ejecta shells. 
On the other hand, the $\pr$ curve of example 6b has no drop at $\tau=7~\mbox{s}$,
and then decays faster than $\pr \propto \rr^{-3}$, just like the example 5b.

All three examples produce a flat plateau phase in the RS X-ray light curves 
while their RS waves sweep up the ejecta shells up until $\tau=7~\mbox{s}$ 
(Figure~\ref{fig:cRX_6a6b4d}). 
The RS wave of example 4d extends the plateau phase slightly as it sweeps 
the remaining ejecta shells (from 7 to 10 s), and then vanishes. 
Thus, its X-ray light curve has $\alpha \sim 3$ (high latitude emission) 
beyond the plateau phase. 
The RS wave of example 6b does not disappear, but becomes weaker and weaker, 
producing its X-ray light curve with $\alpha \sim 2$ beyond the plateau phase.
Lastly, as the RS wave of example 6a continues on the exponential part, 
its X-ray light curve shows the usual $\alpha \sim 1$ phase. 
A steeper decrease between the plateau and $\alpha \sim 1$ phase is 
due to a sudden drop of $\pr$ curve at $\tau=7~\mbox{s}$.

\subsection{Group (6c/6d/6e)}

\begin{figure}
\begin{center}
\includegraphics[width=10cm]{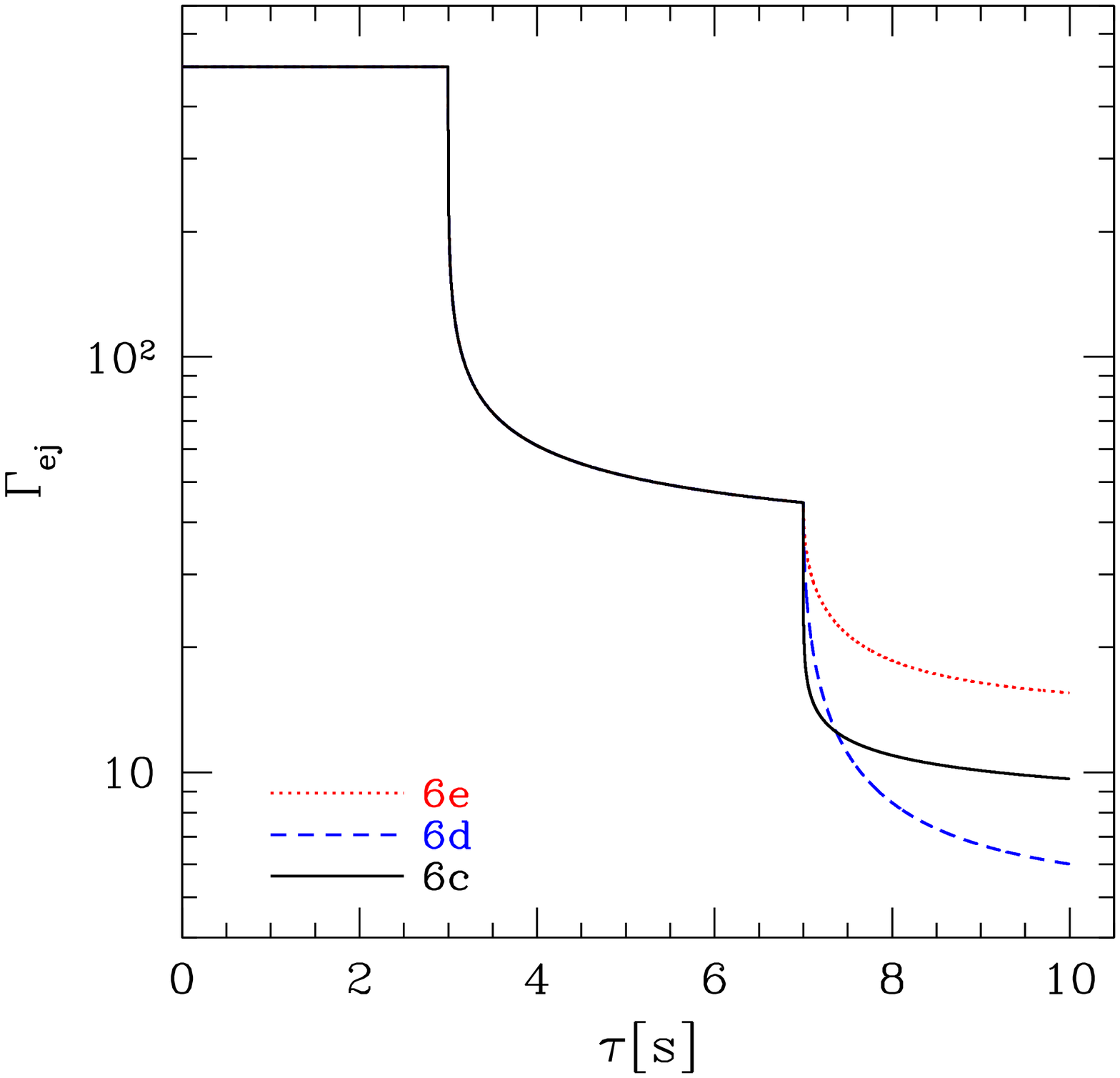}
\caption{
Same as in Figure~\ref{fig:gej_1a1b1c}, but for examples 6c, 6d, and 6e.
} 
\label{fig:gej_6c6d6e} 
\end{center}
\end{figure}

Like the examples 6a and 6b, 
three examples 6c, 6d, and 6e share the same $\Gej$ profile with the example 4d 
up until $\tau=7~\mbox{s}$ (Figure~\ref{fig:gej_6c6d6e}).
From $\tau=7~\mbox{s}$, three examples have a single concave shape 
with various ending values of $\Gej$.

\begin{figure}
\begin{center}
\includegraphics[width=10cm]{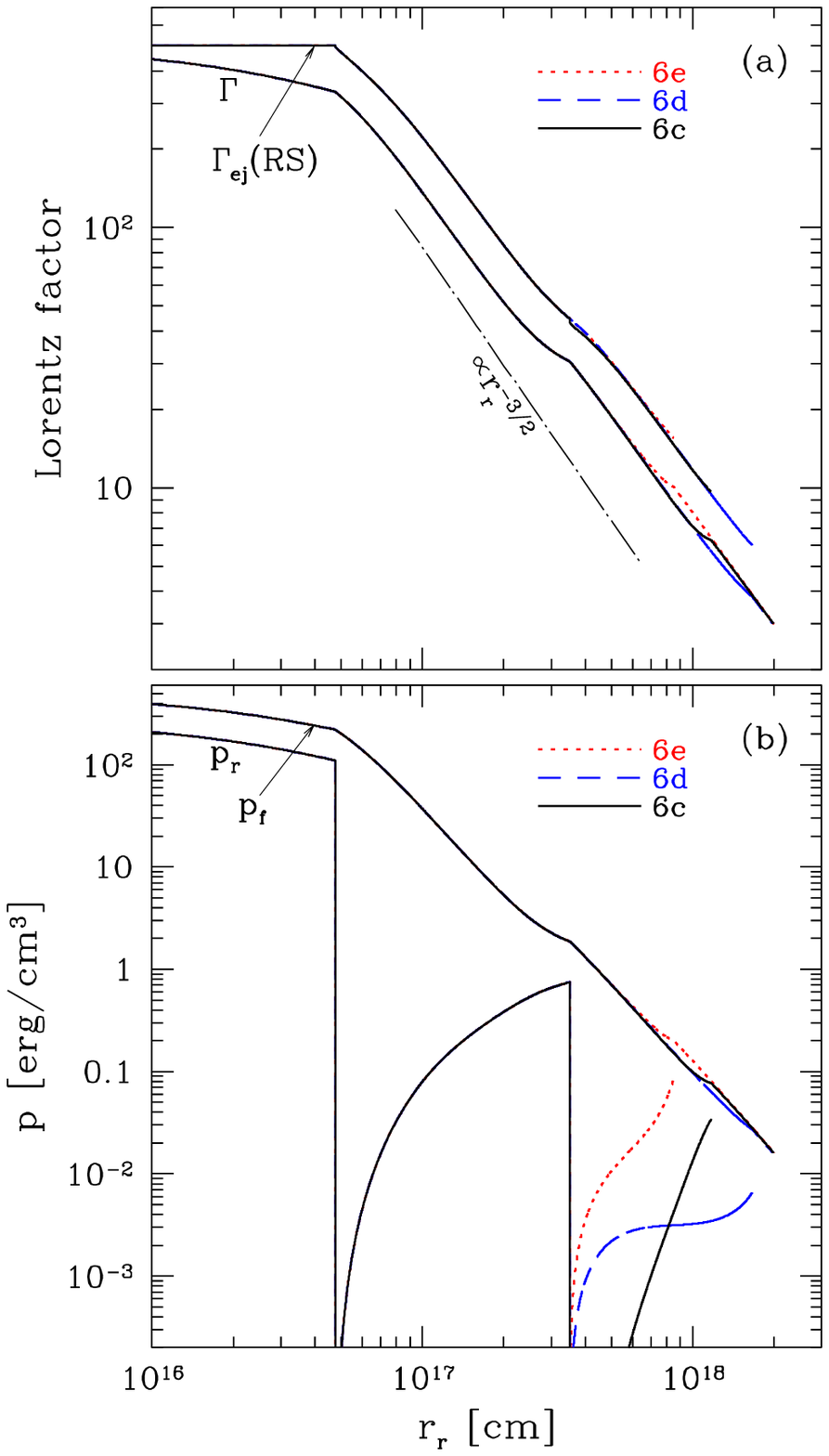}
\caption{
Same as in Figure~\ref{fig:dyn_1a1b1c}, but for examples 6c, 6d, and 6e.
} 
\label{fig:dyn_6c6d6e} 
\end{center}
\end{figure}

The blast wave dynamics are shown in Figure~\ref{fig:dyn_6c6d6e}. 
The RS light curves show a X-ray plateau and then a steep decay ($\alpha \sim 3$), 
followed by three different types of re-brightenings (Figure~\ref{fig:cRX_6c6d6e}).

%--------------------------------------------------------
%
% Section 5 Discussion
%
%--------------------------------------------------------

\section{Discussion} \label{section:discussion}

The $\Gej$ profiles of all 20 examples are shown together in Figure~\ref{fig:gej_all}; 
the line (color) type of each example is the same as previous sections. 
Each example plans on its schedule of spending the same energy budget $E_b=10^{54}~\mbox{ergs}$, 
by shaping its $\Gej$ profile (i.e., its ejecta stratification). 
Depending on this schedule, its $\Gamma$ curve goes through various deceleration path, 
but eventually arrives at the same spot in the $\Gamma$ - $\rr$ plane. 
Indeed, the $\Gamma$ curves of all 20 examples arrive at the same spot 
as is shown in Figure~\ref{fig:dyn_all}.

\begin{figure}
\begin{center}
\includegraphics[width=10cm]{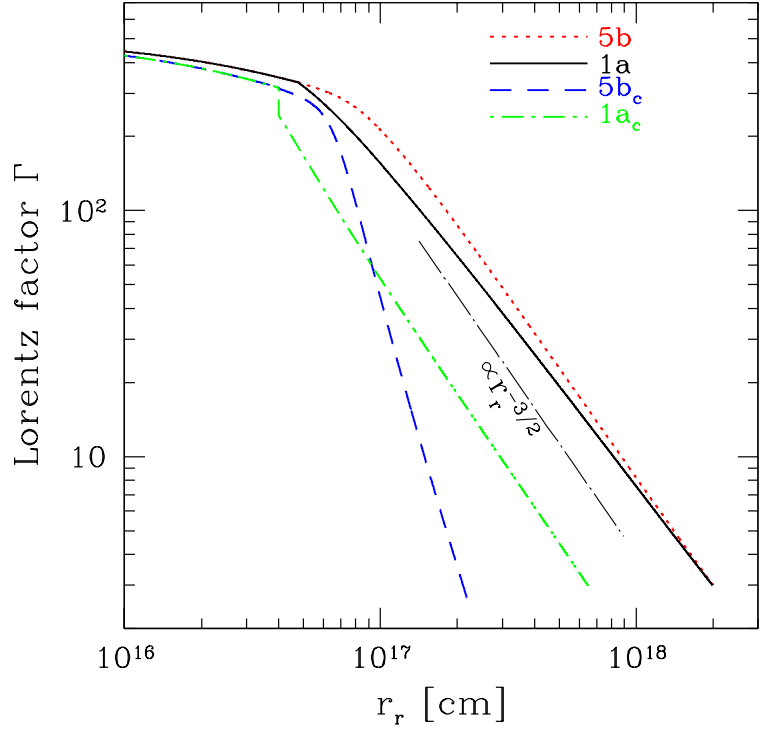}
\caption{
Lorentz factor $\Gamma$ of the blast wave for examples 1a and 5b. 
The $\Gamma$ curves named by 1a and 5b are taken from Figure~\ref{fig:dyn_1a5a5b}.
The $\Gamma$ curves denoted by ${\rm 1a_c}$ and ${\rm 5b_c}$ are found 
for examples 1a and 5b, respectively, 
by making use of a customary pressure balance $\pr=\pf$ across the blast wave. 
} 
\label{fig:dyn_custom} 
\end{center}
\end{figure}

The consistency shown in Figure~\ref{fig:dyn_all} could serve as a ``test'' 
that a blast wave modelling with a long-lived RS needs to pass.
While adopting a customary pressure balance $\pr=\pf$ across the blast wave, 
we find the blast wave dynamics again for examples 1a and 5b. 
The obtained $\Gamma$ curves, denoted by ${\rm 1a_c}$ and ${\rm 5b_c}$, respectively, 
are shown in Figure~\ref{fig:dyn_custom}, together with 
previous $\Gamma$ curves 1a and 5b (taken from Figure~\ref{fig:dyn_1a5a5b}).

The $\Gamma$ curves denoted by ${\rm 1a_c}$ and ${\rm 5b_c}$ clearly show that 
the pressure balance $\pr=\pf$ does not pass the test above. 
Moreover, a sudden drop or deceleration shown in the ${\rm 1a_c}$ curve is 
not physical, since it is a mere consequence of enforcing $\pr=\pf$ on the blast 
when the RS wave encounters a sudden drop in the $\rhoejRS$ curve; recall that 
the example 1a has an abrupt drop in its ejecta density (Section~\ref{section:1a1b1c}). 
The example 5b has no sudden change in its $\rhoejRS$ curve (Section~\ref{section:1a5a5b}).
However, the ${\rm 5b_c}$ curve also reveals a problem with the pressure balance $\pr=\pf$, 
as it decreases faster than $\Gamma \propto \rr^{-3/2}$ 
even though it is meant to describe an adiabatic blast wave for $\rho_1 = \mbox{const.}$ 
An existing long-lived RS would only delay the deceleration of its blast wave, 
as shown by the 1a and 5b curves. 
Hence, this implies that the blast wave dynamics described by the ${\rm 5b_c}$ curve 
does not satisfy the energy conservation. 
In fact, the ${\rm 1a_c}$ curve does not conserve the energy, either; 
its arrival spot is very different from that of the 1a and 5b curves.

Thus, enforcing an equality $\pr=\pf$ or a constant ratio $\pf/\pr = \mbox{const.}$ 
on a blast wave with a long-lived RS would give rise to an incorrect dynamics. 
Furthermore, those methods would not be able to ``capture'' distinctive features 
of the RS dynamics (that are illustrated with 20 examples above), 
since the FS and the RS wave would force each other to follow a similar dynamics.

With a mild energy injection into the blast wave, its deceleration (or the FS dynamics) 
deviates only mildly and smoothly from the solution of BM76, 
as is visible in Figure~\ref{fig:dyn_all}. 
In the meantime, its RS wave responds sensitively to the density flow $\rhoejRS$, 
shaped by various ejecta stratifications, 
and produces fast and strong evolutions in the RS dynamics. 
These distinctive features then show up in the RS light curves, 
so that the RS light curves exhibit more diverse and vigorous behaviors 
than the FS light curves. 
This can be clearly seen in Figure~\ref{fig:cRX_all} 
where we put together the light curves of all 20 examples.

We remark that the diversity and non-trivial features shown in the RS light 
curves (Figure~\ref{fig:cRX_all}) are produced as we simply move through 
the space of ejecta stratifications (Figure~\ref{fig:gej_all}). 
All 20 examples have the same density $\rho_1(r)/m_p=1~\mbox{cm}^{-3}$ 
for the ambient medium, and the same duration $\tau_b=10~\mbox{s}$ and 
isotropic energy $E_b=10^{54}~\mbox{ergs}$ of the burst.
The microphysics parameters are also kept to be the same in all examples. 
Thus, contrary to other proposed scenarios, this diversity does not require 
strong variations on the life time or the energetics of the central engine.

In fact, many features shown in the RS light curves resemble what has been observed 
in the afterglow light curves. In particular, 
a weak bump at the end of the plateau phase, as in example 3c (Figure~\ref{fig:cRX_3a3b3c}), 
has been observed in the X-ray light curve of a few GRB afterglows such as GRB 050502B. 
A steep decay at the end of the plateau, as in example 4d (Figure~\ref{fig:cRX_4a4b4c4d}), 
has been observed in the X-ray light curve of a few GRB afterglows 
such as GRB 060413 and GRB 100508A.
A fast decay at the end of the plateau, followed by a rebrightening, 
as in examples 6c and 6d (Figure~\ref{fig:cRX_6c6d6e}), has been observed in the X-ray 
light curve of a few GRB afterglows such as GRB 100814A, GRB 100418A, and GRB 060607A.

This gives rise to interesting prospects of interpreting puzzling 
afterglow features within the RS scenario. In general, the observed
afterglow emission should be a superposition of the FS and RS
emission components. If the shock parameters of both shocks are
the same, the RS component is usually over-shone by the FS component.
However, since the FS is ultra-relativistic while the RS is mildly relativistic, 
it is possible that the two shocks have different values of $\epsilon_e$ and/or $\epsilon_B$. 
Moreover, as the ejecta may carry a primordial magnetic field from the
central engine, it is reasonable to assume that the RS is more
magnetized than the FS. If the magnetization of the ejecta is
not too high (e.g. the Poynting-flux-to-matter-flux ratio 
$\sigma<0.1$), the magnetization of the ejecta would not 
affect the strength of the RS but would enhance synchrotron
emission from the RS\footnote{In the regime of a higher $\sigma$,
the blast wave dynamics would be affected since the RS shock
jump condition is significantly modified (Zhang \& Kobayashi
2005). The calculations in this paper are relevant when such
an effect is not important.}. 
As a result, the strengthened RS emission
would become comparable or even brighter than the FS emission.
So it is possible that the observed afterglow emission indeed
includes contributions from both the FS and RS.

%--------------------------------------------------------
%
% Section 6 Conclusion
%
%--------------------------------------------------------

\section{Conclusion}\label{section:conclusion}

We have investigated in detail the dynamics and GRB afterglow light curves
for a relativistic blast wave with a long-lived RS. This
long-lived RS is formed in a stratified outflow ejected from 
a central source, which spreads out radially 
and forms various density structures in the flow due to its stratification. 
Due to spreading of Lorentz factor, this ejecta flow gradually catches 
up with the blast wave and adds its kinetic 
energy into the blast, naturally maintaining a long-lived RS wave. 
As a result, this blast wave is not in the self-similar stage, as is
described in the BM76 solution. 
Instead, the blast wave is being continuously pushed by the RS, 
which displays various forms of energy injection scenarios.

\begin{figure}
\begin{center}
\includegraphics[width=10cm]{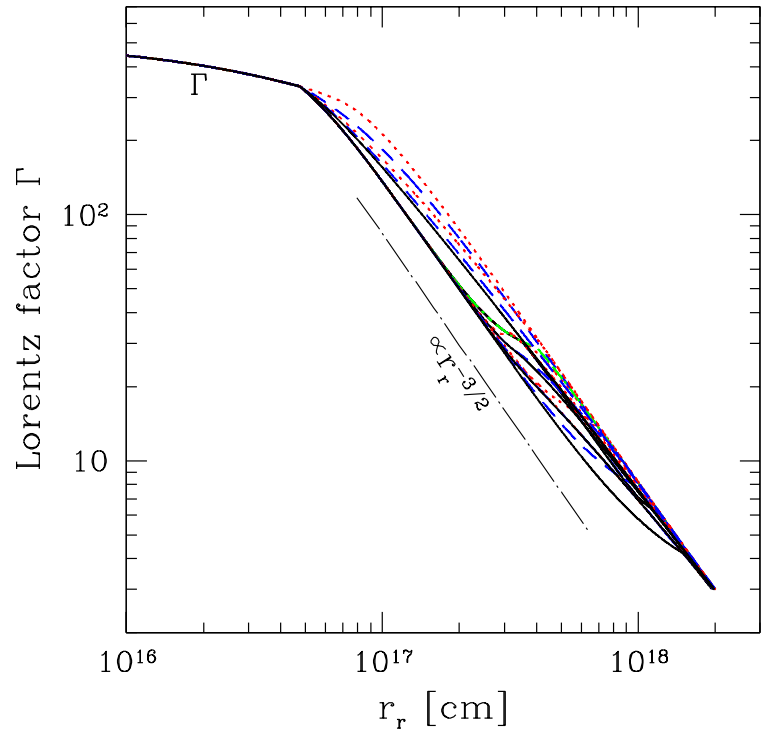}
\caption{
Lorentz factor $\Gamma$ of the blast wave for all 20 examples.
} 
\label{fig:dyn_all} 
\end{center}
\end{figure}

In order to find such dynamics of a blast wave with a long-lived RS, 
we make use of U11 with the mechanical model and perform detailed numerical calculations. 
Investigating a total of 20 different shapes of ejecta stratifications, 
we explain the effects and consequences of radial spreadings on the FS and RS dynamics. 
In particular, we show that there exists a whole new class of the RS dynamics 
with fast and strong evolutions. 
The FS dynamics is also shown to exhibit consistent behaviors for those diverse 
types of energy injections (Figure~\ref{fig:dyn_all}).  
A high accuracy shown in Figure~\ref{fig:dyn_all} indicates that 
we are presenting here a ``precision dynamics'' for the blast waves with a long-lived RS.

Employing a Lagrangian description of the blast wave, 
we perform a sophisticated calculation of afterglows.
In particular, our calculation has (1) a spatial resolution into the blast 
wave region and (2) a pressure profile that smoothly varies over the blast.
For every shell on the blast, we keep track of an evolution of 
(1) the thermodynamic quantities of shocked gas (pressure, energy density, 
adiabatic index, etc) and 
(2) the magnetic field and power-law spectrum of electrons. 
The FS and RS light curves are found by integrating over 
the entire FS and RS shocked regions, respectively, 
while making use of an analytic expression for observed spectral flux 
$\delta F_{\nuobs}^{\rm obs}$, which we derive here in terms of 
an observed frequency $\nuobs$ and observer time $\tobs$ 
(Section~\ref{section:curvature_effect}).

\begin{figure}
\begin{center}
\includegraphics[width=10cm]{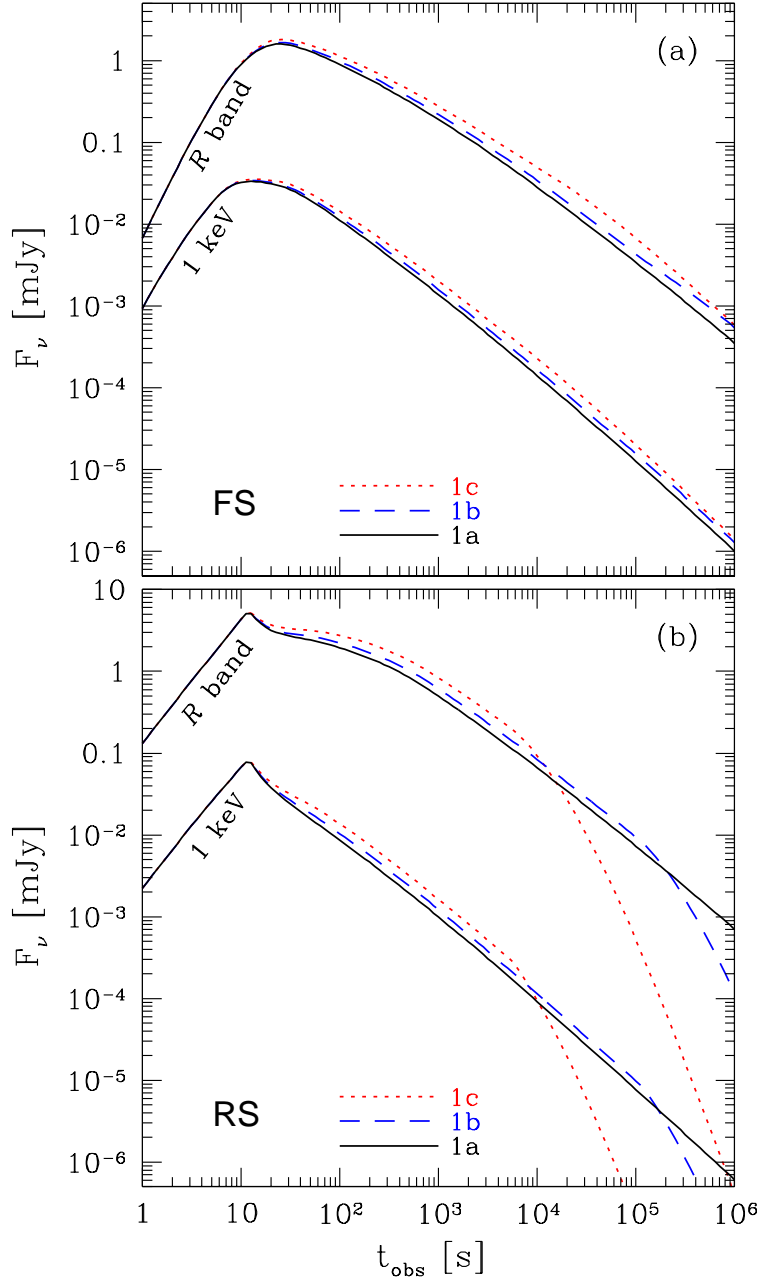}
\caption{
Afterglow light curves for examples 1a, 1b, and 1c. 
The panel (a) shows the FS emissions in X-ray (1 keV) and {\it R} band 
as a function of the observer time $\tobs$. 
The panel (b) shows the RS emissions in X-ray (1 keV) and {\it R} band. 
} 
\label{fig:cRX_1a1b1c}
\end{center}
\end{figure}

\begin{figure}
\begin{center}
\includegraphics[width=10cm]{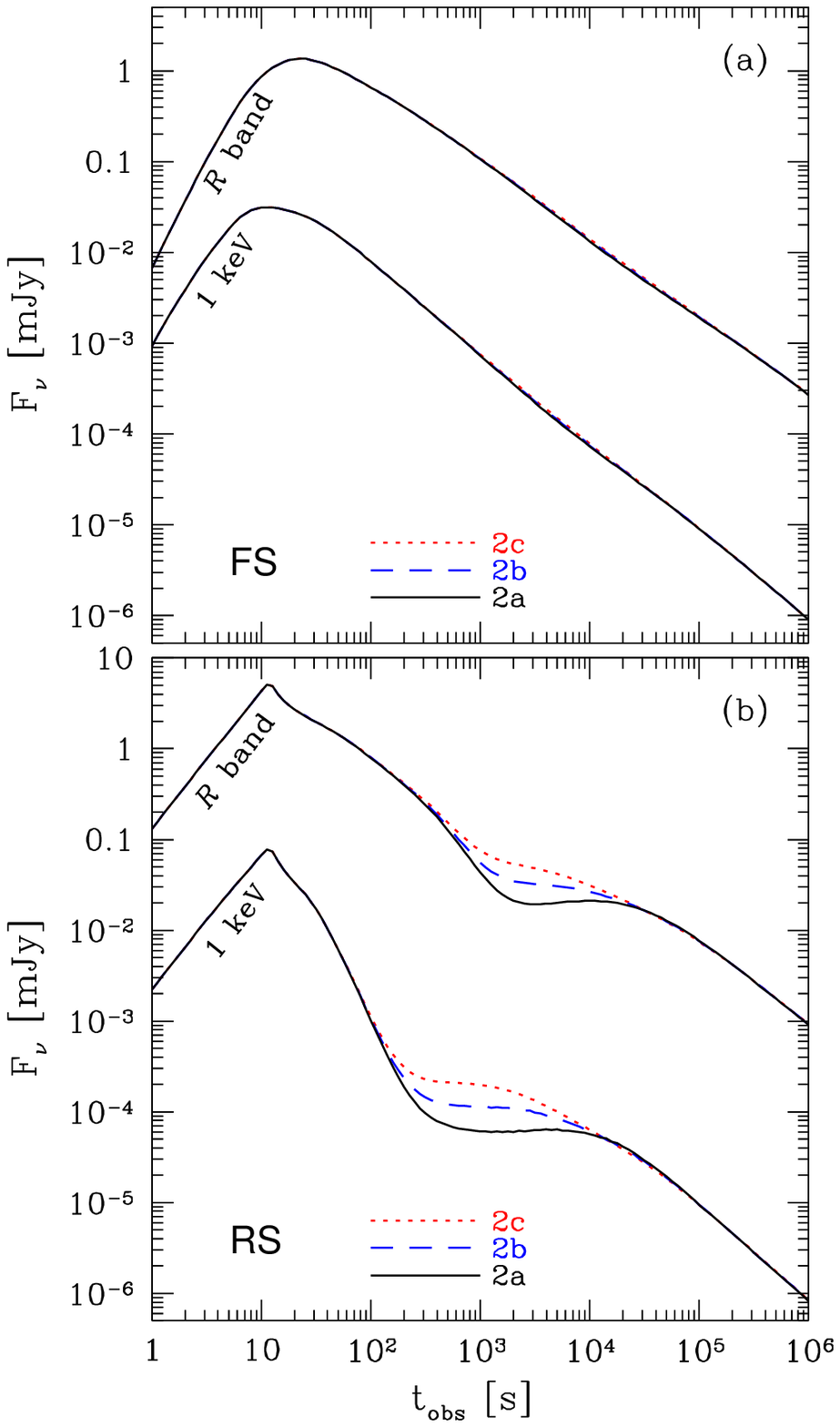}
\caption{
Same as in Figure~\ref{fig:cRX_1a1b1c}, but for examples 2a, 2b, and 2c.
}
\label{fig:cRX_2a2b2c} 
\end{center}
\end{figure}

\begin{figure}
\begin{center}
\includegraphics[width=10cm]{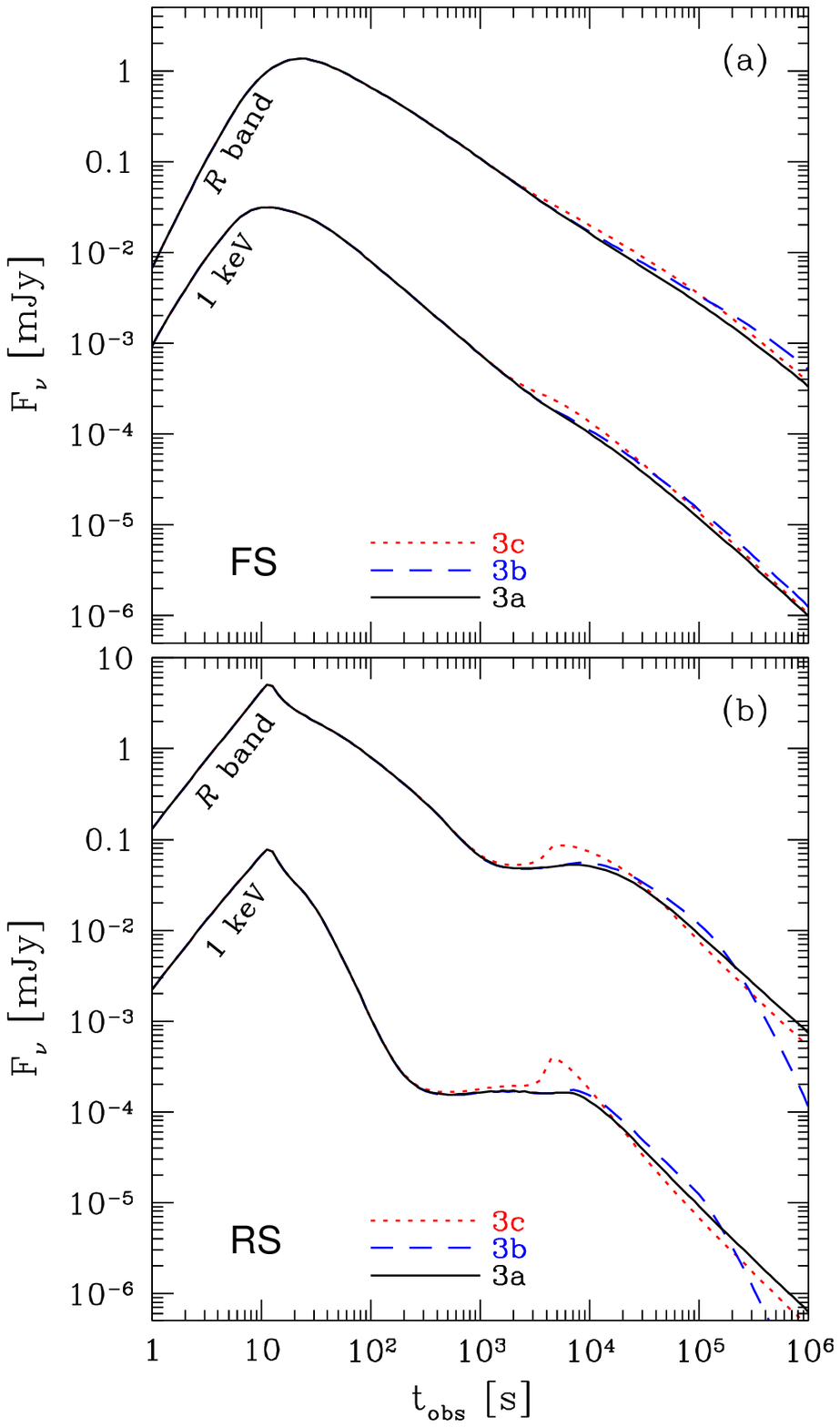}
\caption{
Same as in Figure~\ref{fig:cRX_1a1b1c}, but for examples 3a, 3b, and 3c.
}
\label{fig:cRX_3a3b3c} 
\end{center}
\end{figure}

\begin{figure}
\begin{center}
\includegraphics[width=10cm]{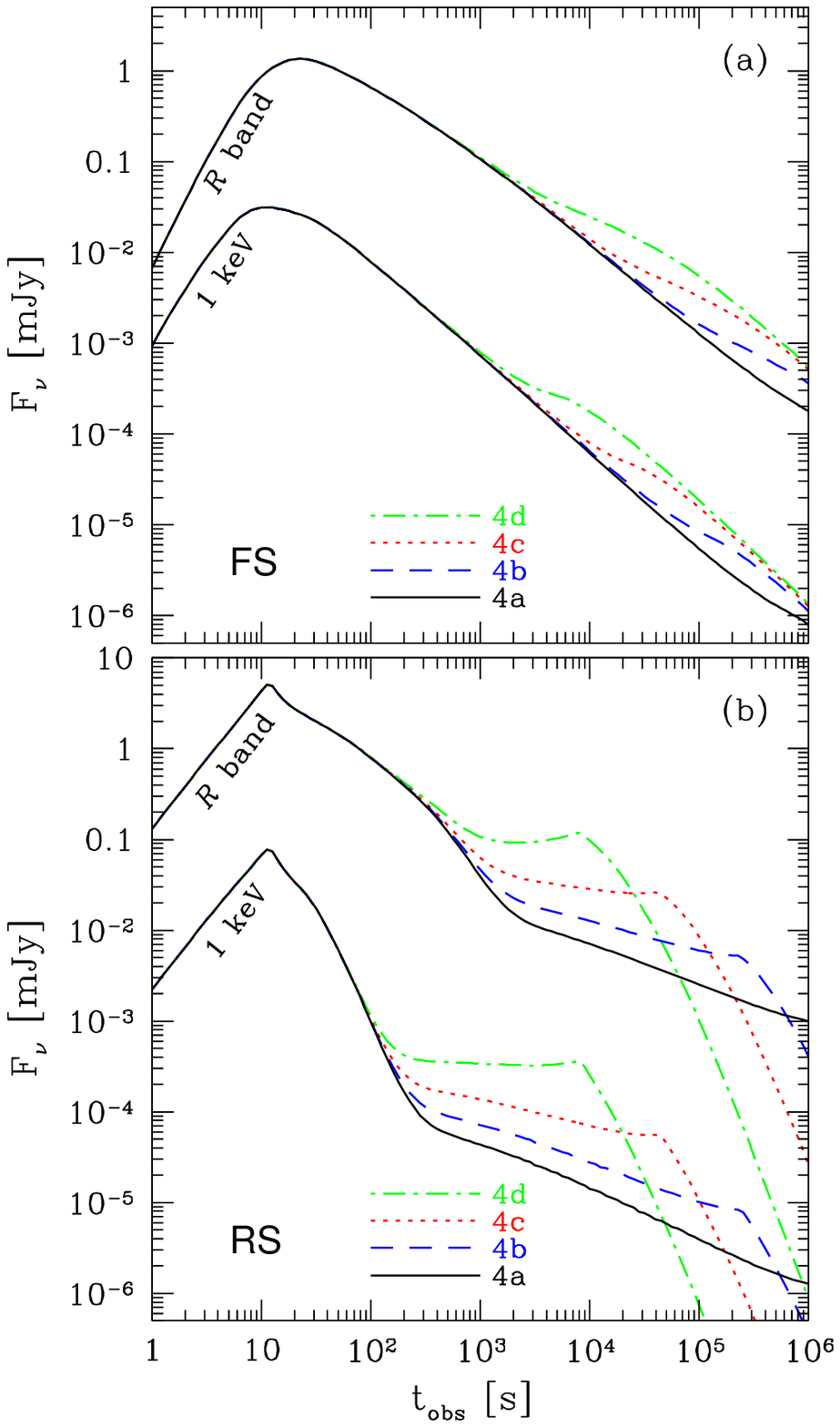}
\caption{
Same as in Figure~\ref{fig:cRX_1a1b1c}, but for examples 4a, 4b, 4c, and 4d.
}
\label{fig:cRX_4a4b4c4d} 
\end{center}
\end{figure}

\begin{figure}
\begin{center}
\includegraphics[width=10cm]{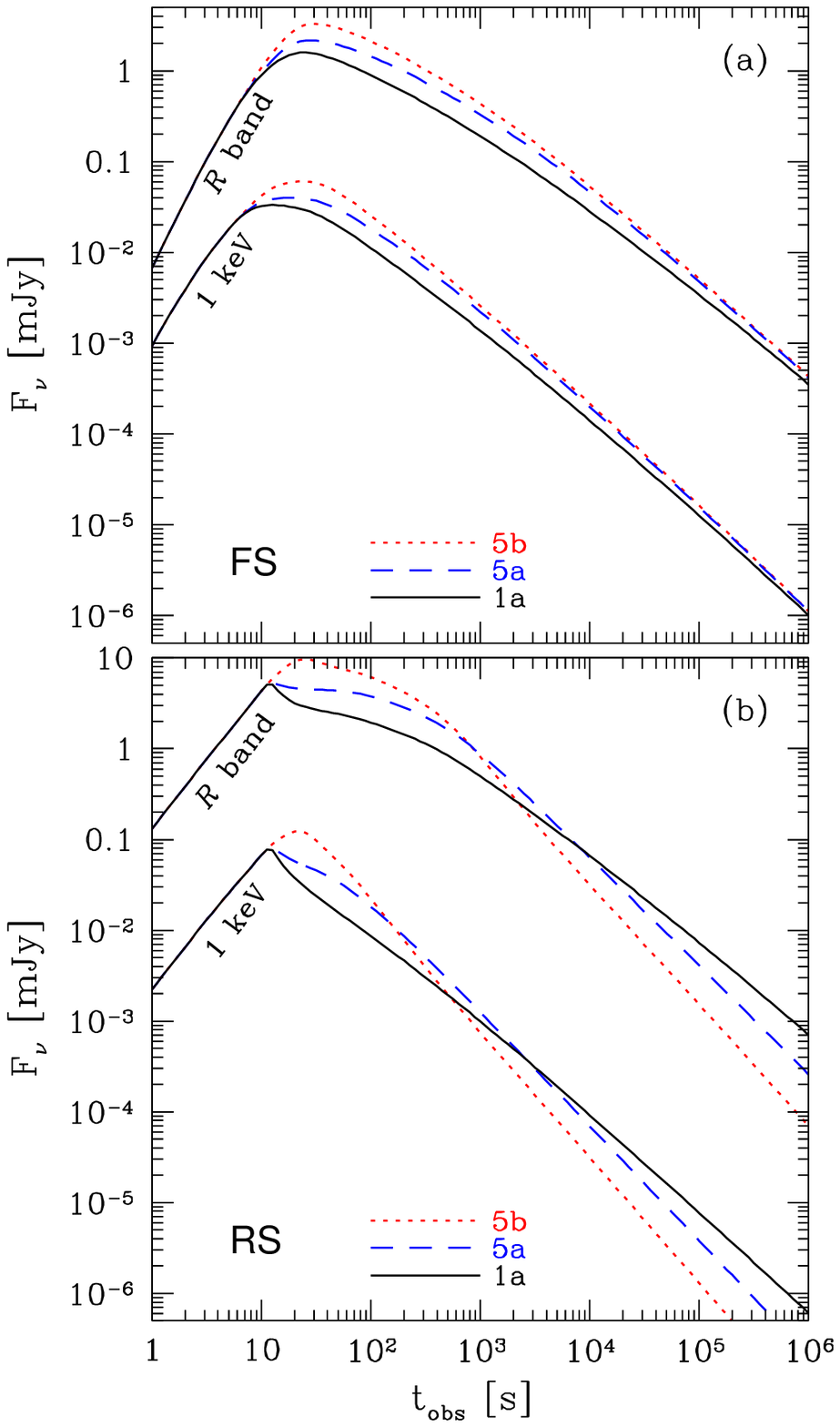}
\caption{
Same as in Figure~\ref{fig:cRX_1a1b1c}, but for examples 1a, 5a, and 5b.
} 
\label{fig:cRX_1a5a5b} 
\end{center}
\end{figure}

\begin{figure}
\begin{center}
\includegraphics[width=10cm]{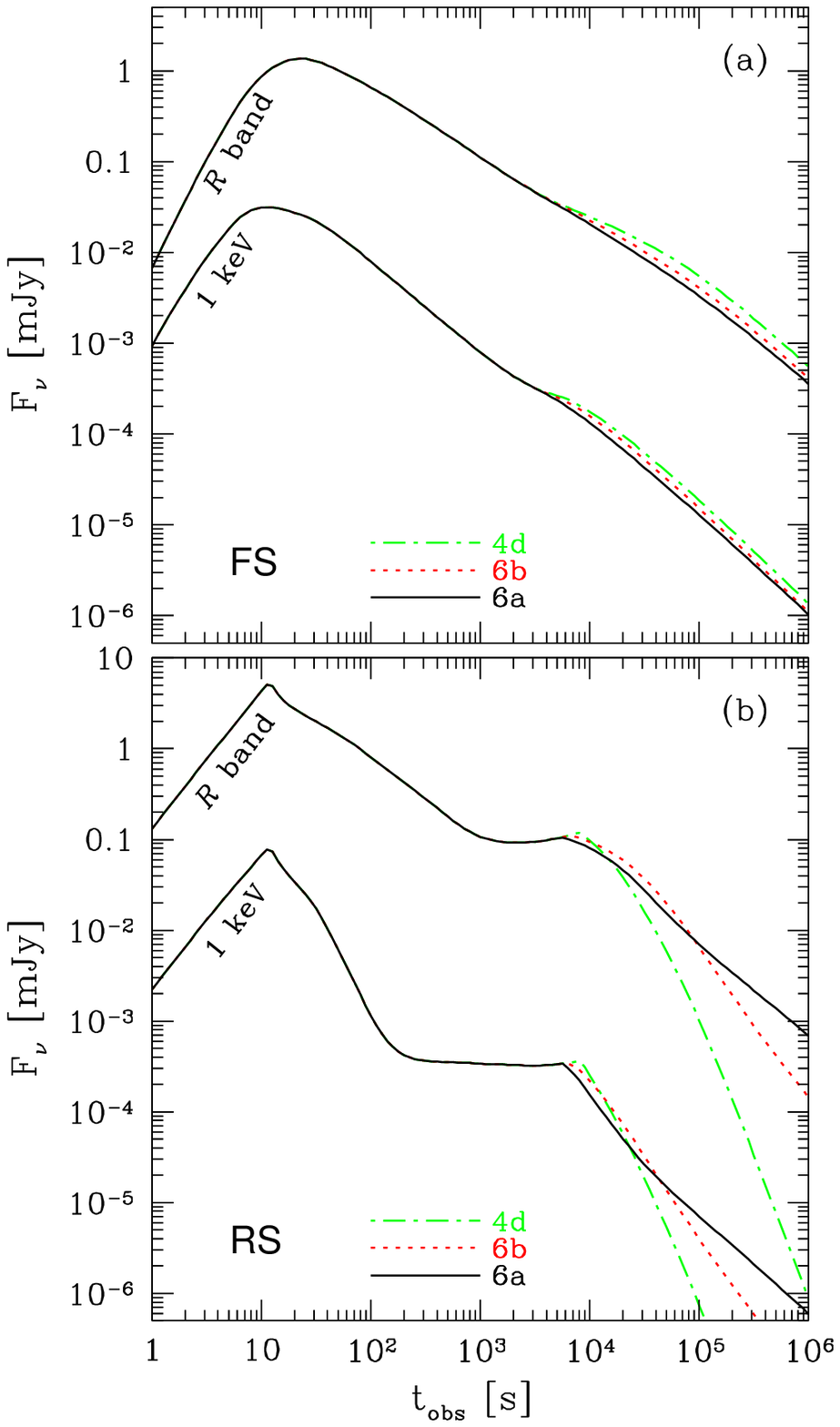}
\caption{
Same as in Figure~\ref{fig:cRX_1a1b1c}, but for examples 6a, 6b, and 4d.
} 
\label{fig:cRX_6a6b4d} 
\end{center}
\end{figure}

\begin{figure}
\begin{center}
\includegraphics[width=10cm]{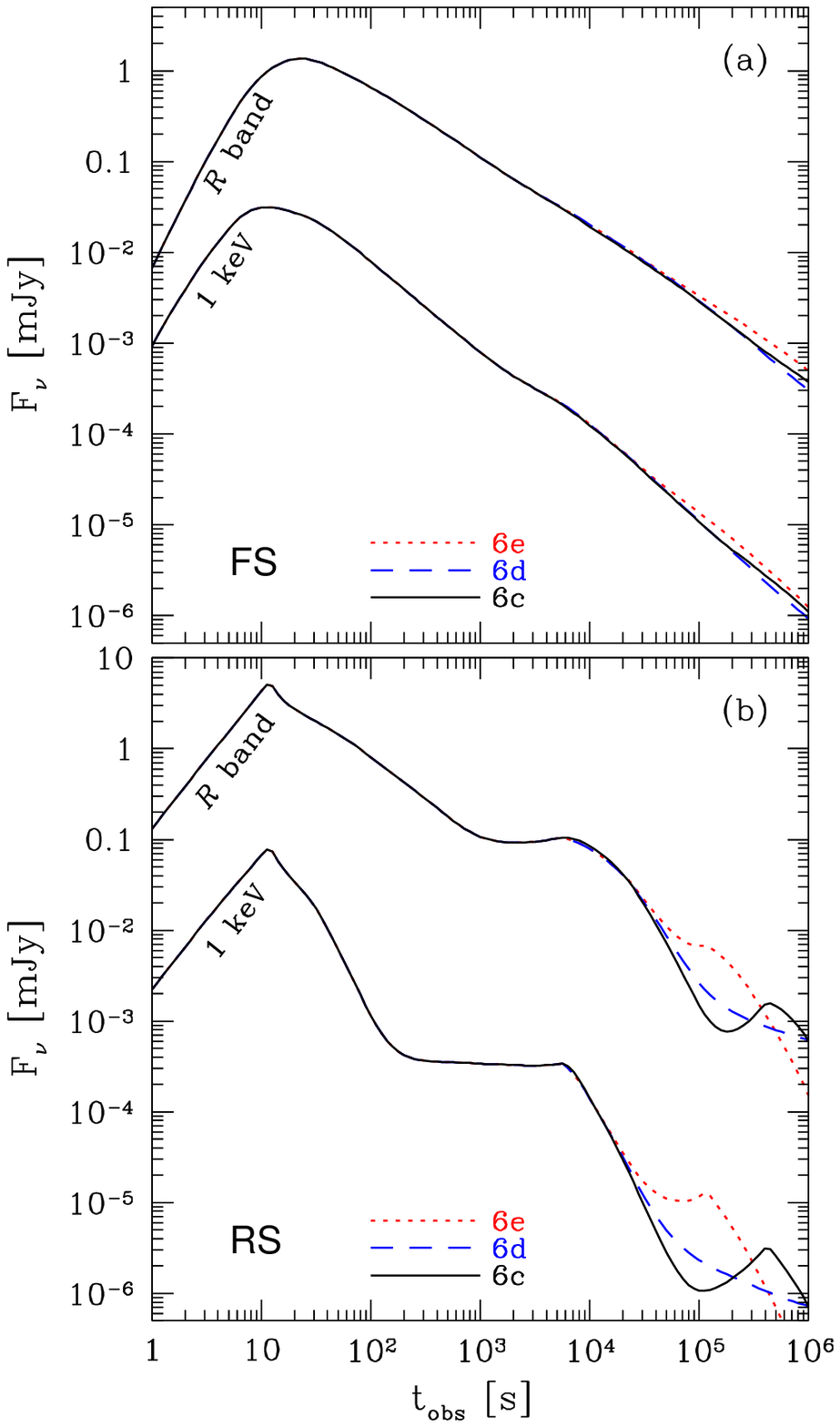}
\caption{
Same as in Figure~\ref{fig:cRX_1a1b1c}, but for examples 6c, 6d, and 6e.
} 
\label{fig:cRX_6c6d6e} 
\end{center}
\end{figure}

\begin{figure}
\begin{center}
\includegraphics[width=10cm]{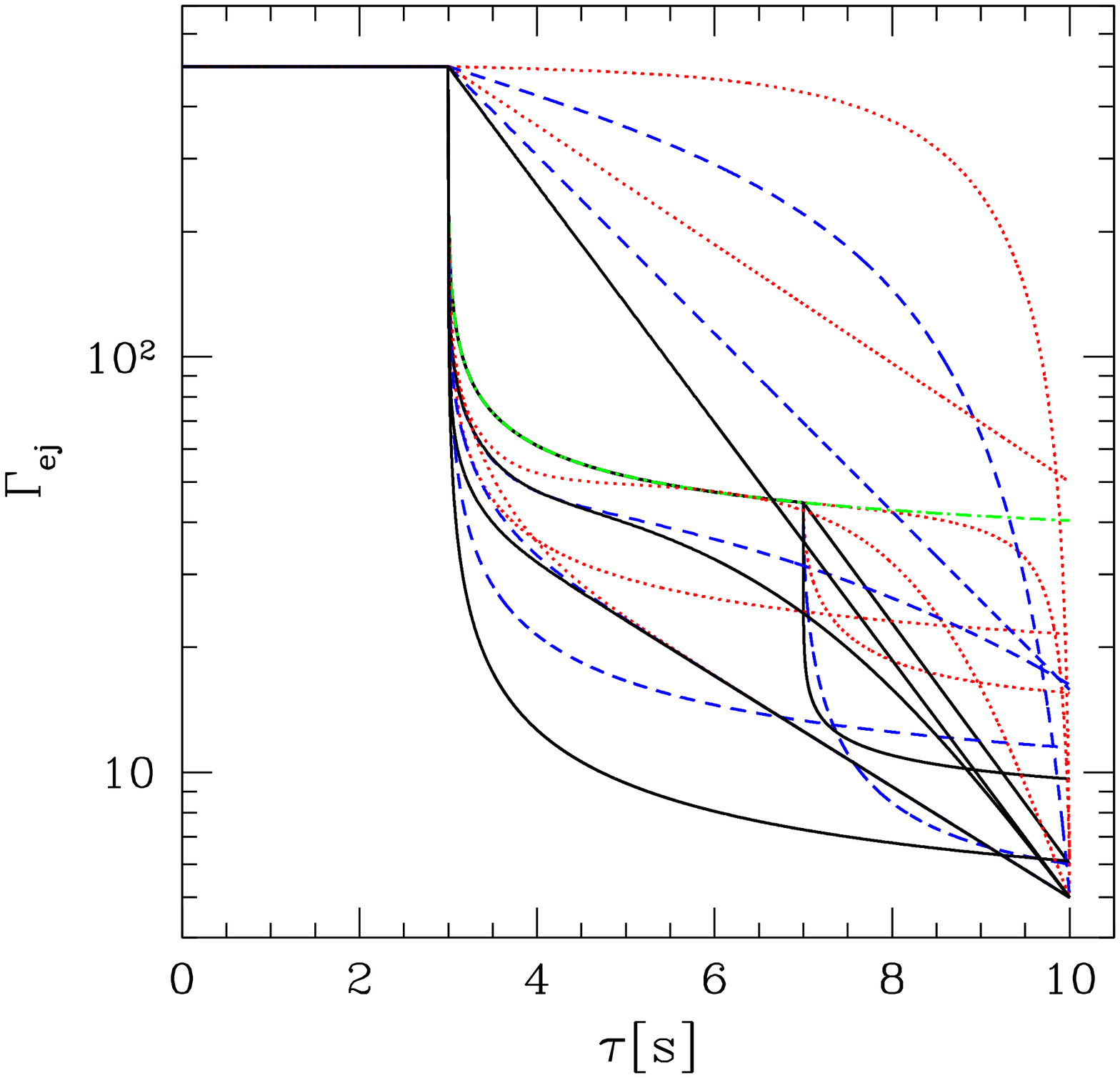}
\caption{
Same as in Figure~\ref{fig:gej_1a1b1c}, but for all 20 examples.
} 
\label{fig:gej_all} 
\end{center}
\end{figure}

\begin{figure}
\begin{center}
\includegraphics[width=10cm]{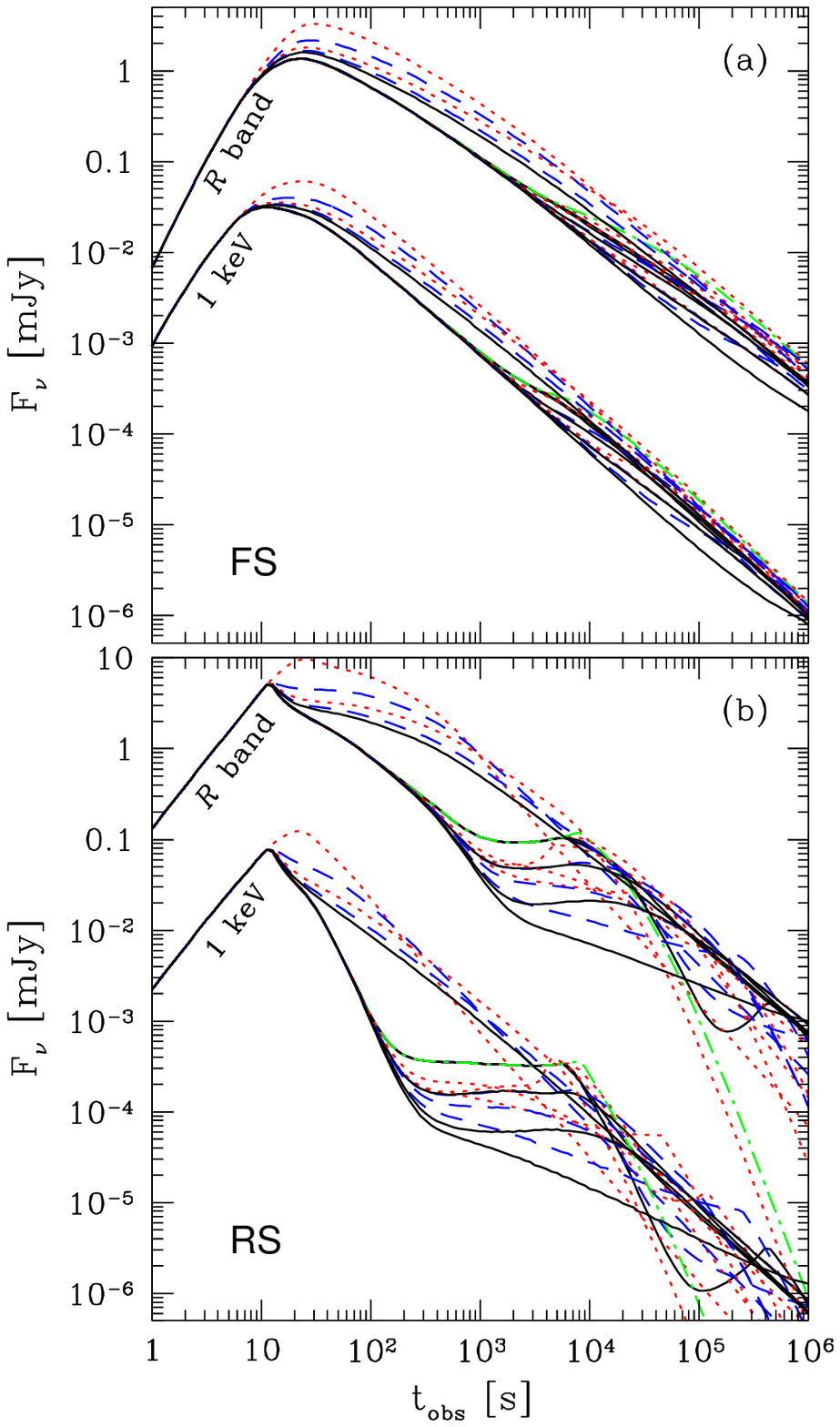}
\caption{
Same as in Figure~\ref{fig:cRX_1a1b1c}, but for all 20 examples.
} 
\label{fig:cRX_all} 
\end{center}
\end{figure}

The resulting afterglow light curves display interesting features.
Since the FS strength mainly depends on the Lorentz factor of the blast, 
the FS light curves do not sensitively depend on the ejecta stratification. 
The strength of the RS, on the other hand, sensitively depends on the ejecta 
density flow $\rhoejRS$ entering the RS, so that rich afterglow features show up 
in the RS light curves. As demonstrated with our 20 examples, the RS light curves 
naturally produce diverse and distinctive features (Figure~\ref{fig:cRX_all}) 
as we move through the space of ejecta stratifications (Figure~\ref{fig:gej_all}). 
In particular, designing proper stratifications in the ejecta, 
the RS light curves reproduce many observed X-ray features, including 
various temporal breaks (Figures~\ref{fig:cRX_1a1b1c} and \ref{fig:cRX_3a3b3c}) and 
decay indices (Figures~\ref{fig:cRX_1a5a5b} and \ref{fig:cRX_6a6b4d}), 
plateaus (Figures~\ref{fig:cRX_2a2b2c}, \ref{fig:cRX_3a3b3c}, and \ref{fig:cRX_4a4b4c4d}), 
steep declines (Figures~\ref{fig:cRX_2a2b2c}, \ref{fig:cRX_4a4b4c4d}, and \ref{fig:cRX_6c6d6e}), 
bumps (Figure~\ref{fig:cRX_3a3b3c}), and 
re-brightenings (Figure~\ref{fig:cRX_6c6d6e}). 
Since the FS and RS could have different efficiency in particle acceleration 
and the GRB ejecta is likely more magnetized than the ambient medium,
it is plausible that the RS emission would become as bright as 
or even brighter than the FS emission. 
Therefore, we believe that the RS could be a strong candidate 
to account for the observed GRB afterglows.

%--------------------------------------------------------
%
% Acknowledgments
%
%--------------------------------------------------------

\acknowledgments

ZLU is grateful to Massimiliano De Pasquale for helpful discussions. 
The authors are grateful to the anonymous referee for useful comments 
to improve the manuscript.
This work was supported by NASA NNX10AD48G, NSF AST-0908362, 
``Research in Paris'' of the Paris City Hall, 
and CRI (RCMST) of MEST/KRF.

%--------------------------------------------------------
%
% References
%
%--------------------------------------------------------

%--------------------------------------------------------
%
% Figures
%
%--------------------------------------------------------

%--------------------------------------------------------
%-------------------------------------------------------- 

%--------------------------------------------------------
%-------------------------------------------------------- 

%--------------------------------------------------------
%-------------------------------------------------------- 

%--------------------------------------------------------
%-------------------------------------------------------- 

%--------------------------------------------------------
%-------------------------------------------------------- 

%--------------------------------------------------------
%-------------------------------------------------------- 

%--------------------------------------------------------
%-------------------------------------------------------- 

%--------------------------------------------------------
%-------------------------------------------------------- 

%--------------------------------------------------------
%-------------------------------------------------------- 

%--------------------------------------------------------
%-------------------------------------------------------- 

%--------------------------------------------------------
%-------------------------------------------------------- 

\clearpage
%--------------------------------------------------------
%-------------------------------------------------------- 

\clearpage
%--------------------------------------------------------
%-------------------------------------------------------- 

\clearpage
%--------------------------------------------------------
%-------------------------------------------------------- 

\clearpage
%--------------------------------------------------------
%-------------------------------------------------------- 

\clearpage
%--------------------------------------------------------
%-------------------------------------------------------- 

\clearpage
%--------------------------------------------------------
%-------------------------------------------------------- 

\clearpage
%--------------------------------------------------------
%-------------------------------------------------------- 

\clearpage
%--------------------------------------------------------
%-------------------------------------------------------- 

\clearpage
%--------------------------------------------------------
%-------------------------------------------------------- 

\clearpage
%--------------------------------------------------------
%-------------------------------------------------------- 

\clearpage
%--------------------------------------------------------
%-------------------------------------------------------- 

\clearpage
%--------------------------------------------------------
%-------------------------------------------------------- 

\clearpage
%--------------------------------------------------------
%-------------------------------------------------------- 

\clearpage
%--------------------------------------------------------
%-------------------------------------------------------- 

\clearpage
%--------------------------------------------------------
%-------------------------------------------------------- 

\clearpage
%--------------------------------------------------------
%-------------------------------------------------------- 

\end{document}